\newcommand{\supportingfigtab}{%
  \setcounter{figure}{0}
  \setcounter{table}{0}
  \renewcommand{\thefigure}{S\arabic{figure}}
  \renewcommand{\thetable}{S\arabic{table}}
}
\newcommand{\supportingequation}{%
  \setcounter{equation}{0}
  \renewcommand{\theequation}{S\arabic{equation}}
}

\documentclass[12pt]{article}
\usepackage{times}
\usepackage[margin=1in,headheight=15pt,headsep=18pt,footskip=28pt]{geometry}
\usepackage{graphicx}
\usepackage{amsmath,amssymb}
\usepackage{authblk}
\usepackage[superscript,nomove]{cite}
\usepackage{booktabs,longtable}
\usepackage{tikz}
\usepackage{listings}
\usepackage{tcolorbox}
\tcbuselibrary{listings, breakable, skins}
\usepackage[table]{xcolor}
\definecolor{idgray}{gray}{0.92}
\usepackage{ragged2e}
\usepackage{microtype}
\usepackage{setspace}
\usepackage{titlesec}
\usepackage{caption}
\usepackage{fancyhdr}
\usepackage{enumitem}
\usepackage[colorlinks=true, linkcolor=blue, citecolor=blue, urlcolor=blue]{hyperref}

\definecolor{prismblue}{RGB}{30,64,175}
\definecolor{prismslate}{RGB}{51,65,85}
\definecolor{prismline}{RGB}{203,213,225}
\definecolor{abstractbg}{RGB}{248,250,252}

\AtBeginDocument{\justifying}

\titleformat{\section}{\normalfont\Large\bfseries\color{black}\RaggedRight}{}{0pt}{}
\titlespacing*{\section}{0pt}{2.2ex plus 0.6ex minus 0.2ex}{1.1ex plus 0.2ex}
\titleformat{\subsection}{\normalfont\large\bfseries\color{black}\RaggedRight}{}{0pt}{}
\titlespacing*{\subsection}{0pt}{1.8ex plus 0.5ex minus 0.2ex}{0.8ex plus 0.2ex}

\setlist{itemsep=0.2em,topsep=0.4em,leftmargin=1.8em}

\newcommand{\publicationnote}{%
  {\footnotesize\itshape Published in \textbf{\textit{npj Computational Materials}}; DOI: \href{https://doi.org/10.1038/s41524-026-02226-3}{10.1038/s41524-026-02226-3}.}%
}

\newtcolorbox{abstractbox}{
  enhanced,
  breakable,
  colback=abstractbg,
  colframe=prismline,
  boxrule=0.6pt,
  arc=2.5pt,
  left=10pt,
  right=10pt,
  top=8pt,
  bottom=8pt,
  before skip=0.7em,
  after skip=0.5em,
  fontupper=\footnotesize,
  borderline west={2.2pt}{0pt}{prismblue}
}

\definecolor{codebg}{RGB}{248,250,252}
\definecolor{codeframe}{RGB}{120,150,190}
\definecolor{codekeyword}{RGB}{0,82,155}
\definecolor{codecomment}{RGB}{88,110,117}
\definecolor{codestring}{RGB}{196,82,44}
\definecolor{codenumber}{RGB}{120,120,120}
\definecolor{codetitle}{RGB}{15,23,42}

\lstdefinestyle{bgocode}{
  language=Python,
  basicstyle={\ttfamily\small\linespread{1.08}\selectfont},
  keywordstyle=\color{codekeyword}\bfseries,
  commentstyle=\color{codecomment}\itshape,
  stringstyle=\color{codestring},
  numberstyle=\scriptsize\color{codenumber}\ttfamily,
  numbers=left,
  numbersep=10pt,
  showstringspaces=false,
  breaklines=true,
  breakatwhitespace=false,
  columns=flexible,
  keepspaces=true,
  tabsize=4,
  upquote=true,
  frame=none,
  xleftmargin=10pt,
  xrightmargin=8pt,
  aboveskip=6pt,
  belowskip=4pt
}
\newtcolorbox{apicodebox}[1]{
  enhanced,
  breakable,
  colback=codebg,
  colframe=prismblue,
  coltitle=white,
  fonttitle=\bfseries\footnotesize,
  title={#1},
  attach boxed title to top left={xshift=8pt,yshift=-2.5mm},
  boxed title style={
    colback=prismblue,
    colframe=prismblue,
    boxrule=0pt,
    arc=2pt,
    left=6pt,
    right=6pt,
    top=2pt,
    bottom=2pt
  },
  boxrule=0.45pt,
  arc=3pt,
  left=10pt,
  right=10pt,
  top=12pt,
  bottom=8pt,
  before skip=0.9em,
  after skip=0.8em,
  borderline west={2.4pt}{0pt}{prismblue},
  drop fuzzy shadow=black!12
}

\title{\vspace{-3.0em}\publicationnote\\[-0.45em]
{\color{prismblue}\rule{0.72\textwidth}{1.1pt}}\\[0.7em]
\textbf{Bgolearn: a unified Bayesian optimization framework for accelerating materials discovery}}

\author[1,2]{Bin Cao}
\author[3,4,$*$]{Jie Xiong}
\author[5]{Jiaxuan Ma}
\author[3]{Yuan Tian}
\author[3]{Yirui Hu}
\author[6]{Mengwei He}
\author[1]{Longhan Zhang}
\author[7]{Jiayu Wang}
\author[8,$*$]{Jian Hui}
\author[7]{Li Liu}
\author[9]{Dezhen Xue}
\author[9,10,$*$]{Turab Lookman}
\author[11,$*$]{Jun Wang}
\author[1,2,$*$]{Tong-Yi Zhang}

\affil[1]{Guangzhou Municipal Key Laboratory of Materials Informatics, Advanced Materials Thrust, 
The Hong Kong University of Science and Technology (Guangzhou), Guangzhou, China}

\affil[2]{Department of Physics, City University of Hong Kong, Hong Kong SAR, China}

\affil[3]{Materials Genome Institute, Shanghai University, Shanghai, China}

\affil[4]{State Key Laboratory of Materials for Advanced Nuclear Energy, Shanghai University, Shanghai, China}

\affil[5]{School of Materials Science and Engineering, Shanghai Jiao Tong University, Shanghai, China}

\affil[6]{School of Aerospace, Mechanical and Mechatronic Engineering, School of Computer Science, The University of Sydney, Sydney, NSW, Australia}

\affil[7]{School of Materials Science and Engineering, Harbin Institute of Technology (Shenzhen), Shenzhen, China}

\affil[8]{Suzhou Laboratory, Suzhou, China}

\affil[9]{State Key Laboratory for Mechanical Behavior of Materials, Xi’an Jiaotong University, Xi’an, China}

\affil[10]{AiMaterials Research, Santa Fe, NM, USA}

\affil[11]{University College London, London, UK}

\affil[$*$]{Corresponding authors: xiongjie@shu.edu.cn; huij@szlab.ac.cn; turablookman@gmail.com; jun.wang@cs.ucl.ac.uk; mezhangt@hkust-gz.edu.cn}

\date{}

\begin{document}

\maketitle

\begin{abstractbox}
\noindent{\normalsize\bfseries Abstract}\par\vspace{0.25em}
\noindent
Efficient exploration of vast compositional and processing spaces remains a major challenge in accelerated materials discovery. Bayesian optimization (BO) provides a principled approach to identify optimal materials with minimal experimentation, but its adoption has been limited by implementation complexity and a lack of domain-specific tools. Here, we present Bgolearn, a versatile Python framework that brings BO to materials research through intuitive interfaces, robust algorithms, and materials-focused workflows. Bgolearn supports single- and multi-objective optimization, multiple acquisition strategies, diverse surrogate models, and uncertainty quantification, enabling effective navigation of complex design spaces. Benchmark studies show that Bgolearn reduces experimental effort by 40–60\% compared with random search, grid search, and genetic algorithms, while achieving comparable or superior solution quality. Its effectiveness is demonstrated across case studies, including the discovery of maximum-elastic-modulus triply periodic minimal surface structures, ultra-high-hardness high-entropy alloys, and high-strength, high-ductility medium-Mn steels, and is further supported by numerous publications. With a modular architecture that integrates seamlessly into existing materials workflows and a graphical interface (BgoFace) that removes programming barriers, Bgolearn establishes a practical, reliable platform for Bayesian optimization in materials science. The software is openly available at \url{https://github.com/Bin-Cao/Bgolearn}.
\end{abstractbox}

\vspace{0.1em}
\noindent{\footnotesize\color{prismslate}\textbf{Keywords:} Bayesian optimization; Bgolearn; Multi-objective; Modular architecture; Materials discovery}
\newpage

\section*{Introduction}

The accelerated discovery of advanced materials is critical for addressing global challenges in energy, sustainability, and technology\cite{lookman2016information}. However, traditional materials development relies on intuition-guided experimentation or exhaustive design-of-experiments approaches that scale poorly with problem complexity. A typical materials optimization problem involves 5-15 design variables spanning composition ratios, processing temperatures, annealing times, and processing atmospheres, with each experimental iteration requiring days to weeks and costing hundreds to thousands of dollars \cite{lookman2019active}. This combinatorial explosion limits researchers to exploring less than 0.1\% of feasible design spaces, leaving vast regions of potentially superior materials undiscovered\cite{ling2017high}.

Bayesian optimization (BO) provides a principled framework for accelerating experimental discovery by constructing probabilistic surrogate models from limited data and using acquisition functions to balance exploration of uncertain regions with exploitation of promising candidates~\cite{roussel2024bayesian,li2024sequential,sabanza2025best}. By intelligently selecting the most informative next experiments, BO can drastically reduce the number of required iterations compared to random or grid-based search methods.
Recent studies have demonstrated BO's growing impact in materials science. For instance, a benchmarking study across multiple experimental domains showed that BO significantly outperforms random or grid search, particularly when surrogate models and hyperparameters are carefully optimized~\cite{liang2021benchmarking}. Jang \textit{et al.}~\cite{jang2025active} applied BO with sparse experimental data (80 out of 16,206 samples) to efficiently identify three new high-entropy chalcogenides (HECs) exhibiting exceptional thermoelectric performance ($zT > 2$). Similarly, Tian \textit{et al.}~\cite{tian2025materials} proposed a ``target-oriented'' BO framework designed not merely to maximize or minimize a property but to tune it toward a desired target value, for example, achieving a transformation temperature in a shape-memory alloy within $2.66~^{\circ}\mathrm{C}$ of the target after only three experiments.
Other recent works~\cite{gonzalez2024survey} have extended BO to discrete, high-dimensional search spaces, such as those encountered in chemical and biological material design, demonstrating superior efficiency compared with state-of-the-art methods. Overall, the successful application of BO to materials optimization and discovery has been repeatedly validated across diverse materials systems, as evidenced by a series of publications~\cite{jablonka2021bias,sun2025artificial,shields2021bayesian,linden2025increasing,li2024sequential,ma2024mlmd,yu2026efficient,khan2022antbo,hase2018phoenics,burger2020mobile,shields2021bayesian,wu2024race,macleod2020self}, which are not listed here individually.

These developments suggest that the central challenge is no longer the value of BO for materials science, but rather how to make BO workflows accessible, reproducible, and readily deployable by non-specialists for routine materials optimization~\cite{yu2026efficient}. Bgolearn was developed in this context, not to claim conceptual novelty for BO itself, but to bridge the gap between the rapid expansion of BO applications and the practical software infrastructure needed for everyday materials research. Although algorithmically general-purpose, Bgolearn is particularly well suited to materials problems because it supports small and costly datasets, mixed composition–processing variables, multi-objective trade-offs, experiment-in-the-loop decision making, and graphical user interface (GUI) assisted operation for users without extensive programming expertise.

In many cases, the primary obstacle is not the lack of available BO methods, but the difficulty of deploying them in ways that are transparent, flexible, and accessible to domain researchers. Several challenges are particularly relevant. First, widely used BO libraries such as BoTorch~\cite{balandat2020botorch}, AE~\cite{bakshy2018ae}, PHYSBO~\cite{PHYSBO-paper2022}, and GPyOpt~\cite{gpyopt2016} provide powerful algorithmic foundations, yet their effective application often requires substantial machine-learning expertise, particularly for selecting surrogate models, tuning hyperparameters, and diagnosing convergence behavior. Second, although multi-objective optimization is central to many materials-design problems, practical workflows for exploring property trade-offs are often less straightforward than those for single-objective optimization. This challenge is especially pronounced in tasks involving competing objectives, such as strength versus ductility, conductivity versus cost, or performance versus processability, where rigorous Pareto-front exploration using expected hypervolume improvement (EHVI)~\cite{emmerich2018tutorial} or q-noisy expected hypervolume improvement (qNEHVI)~\cite{daulton2021parallel} is highly desirable but not always readily accessible in routine research practice (see Supporting Materials Section~\ref{Algorithm} for details). Third, standard GP surrogate models, while appealing due to their principled probabilistic formulation, become computationally expensive as the number of observations grows and are often less convenient for design spaces containing discrete or categorical variables, which frequently arise in materials formulations. Fourth, scalable alternatives such as random forests or gradient boosting methods are often attractive in practice, but their uncertainty estimates are generally less standardized than GP posteriors, complicating their integration into uncertainty-aware optimization workflows.  

\begin{figure}[h]
    \centering
    \includegraphics[width=0.95\columnwidth]{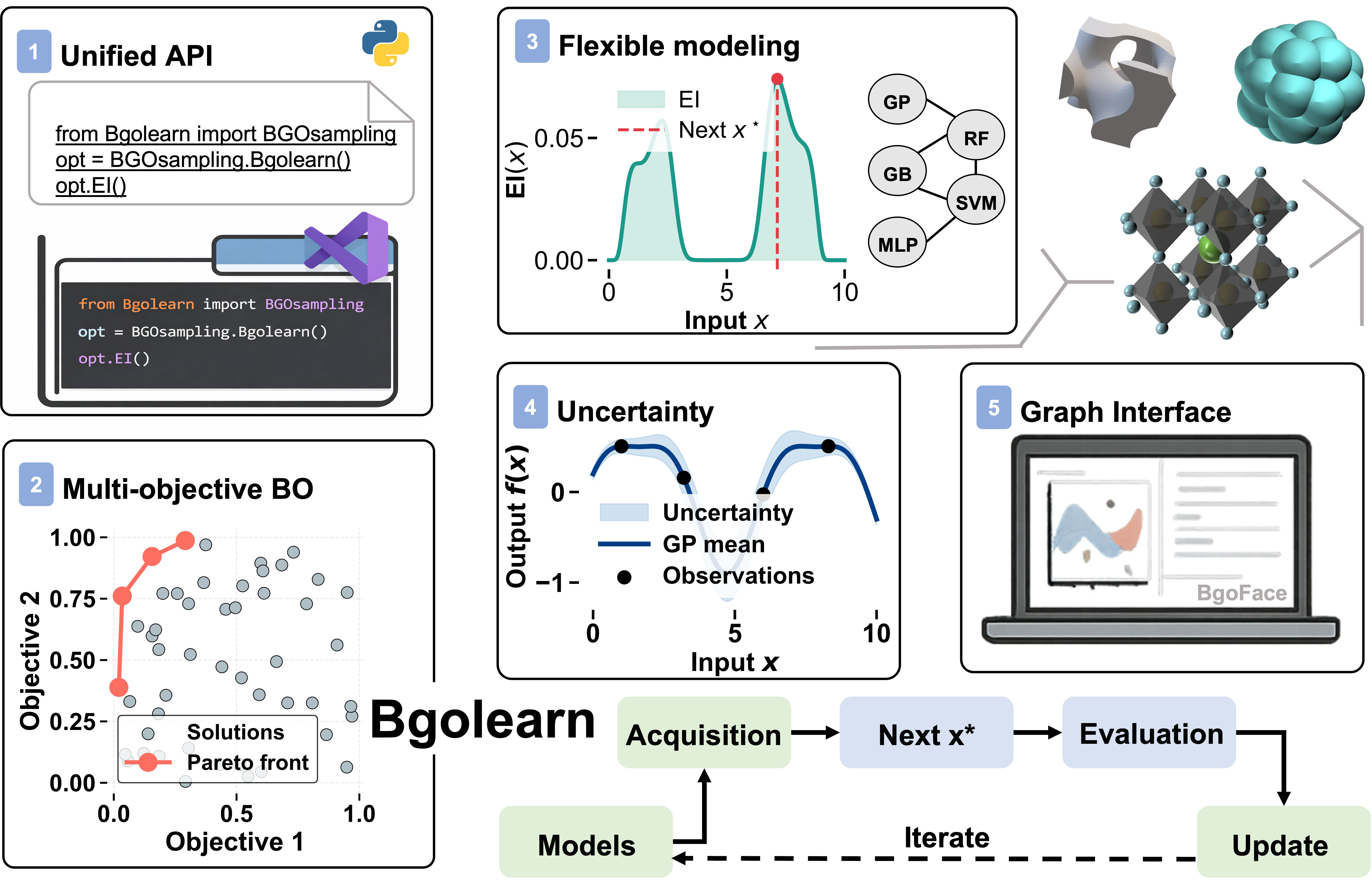}
    \caption{\textbf{The components and workflow of Bgolearn for materials discovery.} Bgolearn integrates data preparation, surrogate modeling, acquisition-function selection, candidate recommendation, and feedback into a unified Bayesian optimization workflow, supporting both single- and multi-objective materials optimization through code-based or GUI-assisted interfaces.} 
    \label{fig:bgolearn_wf}
\end{figure}

Motivated by these practical considerations, we developed Bgolearn as a production-ready BO toolkit that preserves algorithmic rigor while making Bayesian optimization more accessible and usable for materials researchers. The framework introduces five key features (Fig.~\ref{fig:bgolearn_wf}). First, it provides a unified, materials-oriented API that reduces complex BO workflows to as few as 3–5 lines of code while retaining full flexibility and customizability. Second, Bgolearn offers comprehensive support for multi-objective Bayesian optimization (MOBO), including expected hypervolume improvement, q-noisy expected hypervolume improvement, multi-objective probability of improvement (MO-PI), and multi-objective upper confidence bound (MO-UCB) acquisition functions, enabling principled exploration of multi-property trade-offs. Third, the framework supports flexible surrogate modeling through Gaussian processes, random forests (RF), gradient boosting (GB), support vector regression (SVR), and neural networks, together with automatic model selection based on cross-validation performance. Fourth, Bgolearn incorporates bootstrap-based uncertainty quantification for non-GP models, enabling scalable MOBO while alleviating the computational bottlenecks associated with GP surrogates. Finally, the framework includes BgoFace, a GUI designed to lower programming barriers while automatically generating equivalent Python code to enhance reproducibility and transparency. 

Since its first open-source release in 2022, Bgolearn has been downloaded more than 125{,}000 times worldwide \cite{bgolearn-downloads} and applied to a broad range of materials optimization problems, including lead-free solder alloy design~\cite{cao2024active}, nanozyme optimization~\cite{li2025optimize,li2025machine}, magnesium alloy development~\cite{zhang2025accelerated}, electromagnetic metamaterials design~\cite{liu2025bayesian}, foam-agent formulation~\cite{wang2025active}, acidic oxygen evolution reaction (OER) catalyst discovery~\cite{cao2025spatial}, and self-driving characterization platforms~\cite{li2025self}. These studies demonstrate the growing need for BO infrastructures that are not only algorithmically robust, but also practical for day-to-day experimental and computational materials workflows. In the following sections, we first summarize representative applications of Bgolearn and then describe the framework architecture, implementation, and benchmarking results. We further demonstrate its performance through real-world materials-discovery case studies. Detailed mathematical formulations and implementation details are provided in the Methods section and Supporting Materials.

\newpage
\section*{Results}
\subsection*{Bgolearn is driving faster materials discovery.} Bgolearn has already demonstrated practical impact across a wide range of materials discovery and optimization problems, enabling data-efficient exploration of complex design spaces while substantially reducing experimental cost and iteration time~\cite{cao2024active,li2025optimize,li2025machine,zhang2025accelerated,liu2025bayesian,wang2025active,li2025self,cao2025spatial}. Rather than being limited to benchmark studies, Bgolearn has been integrated into real experimental workflows spanning metallic alloys, functional materials, catalysis, industrial formulations, and autonomous laboratories.

Representative studies show that Bgolearn can rapidly identify high-performing candidates with only a small number of experimental iterations. In alloy design, it has been used to optimize both Sn--Ag--Cu lead-free solders and Mg--Ca--Zn systems, successfully balancing strength and ductility while navigating high-dimensional composition spaces~\cite{cao2024active,zhang2025accelerated}. In functional materials research, Bgolearn has accelerated the discovery of high-quantum-yield nanozyme systems and enabled efficient optimization of broadband polarization-insensitive metasurfaces~\cite{li2025optimize,li2025machine,liu2025bayesian}. Beyond property optimization alone, the framework has also supported adaptive exploration of catalyst activity and stability trade-offs, reducing the need for costly durability evaluations in acidic OER catalyst discovery~\cite{cao2025spatial}.

Importantly, Bgolearn is increasingly being deployed in realistic engineering and autonomous experimentation environments. Active-learning-guided optimization has been applied to industrial foam-agent formulations for shield tunneling applications~\cite{wang2025active}, while recent work integrated Bgolearn into a self-driving laboratory for graphene nanoribbon synthesis under ultrahigh vacuum conditions~\cite{li2025self}. Coupled with robotic experimentation and in-situ characterization, the system achieved target nanoribbon morphologies within only twelve closed-loop experimental cycles.

\subsection*{Bgolearn adopts a modular package architecture.} 
\begin{figure}[h]
    \centering
    \includegraphics[width=0.95\columnwidth]{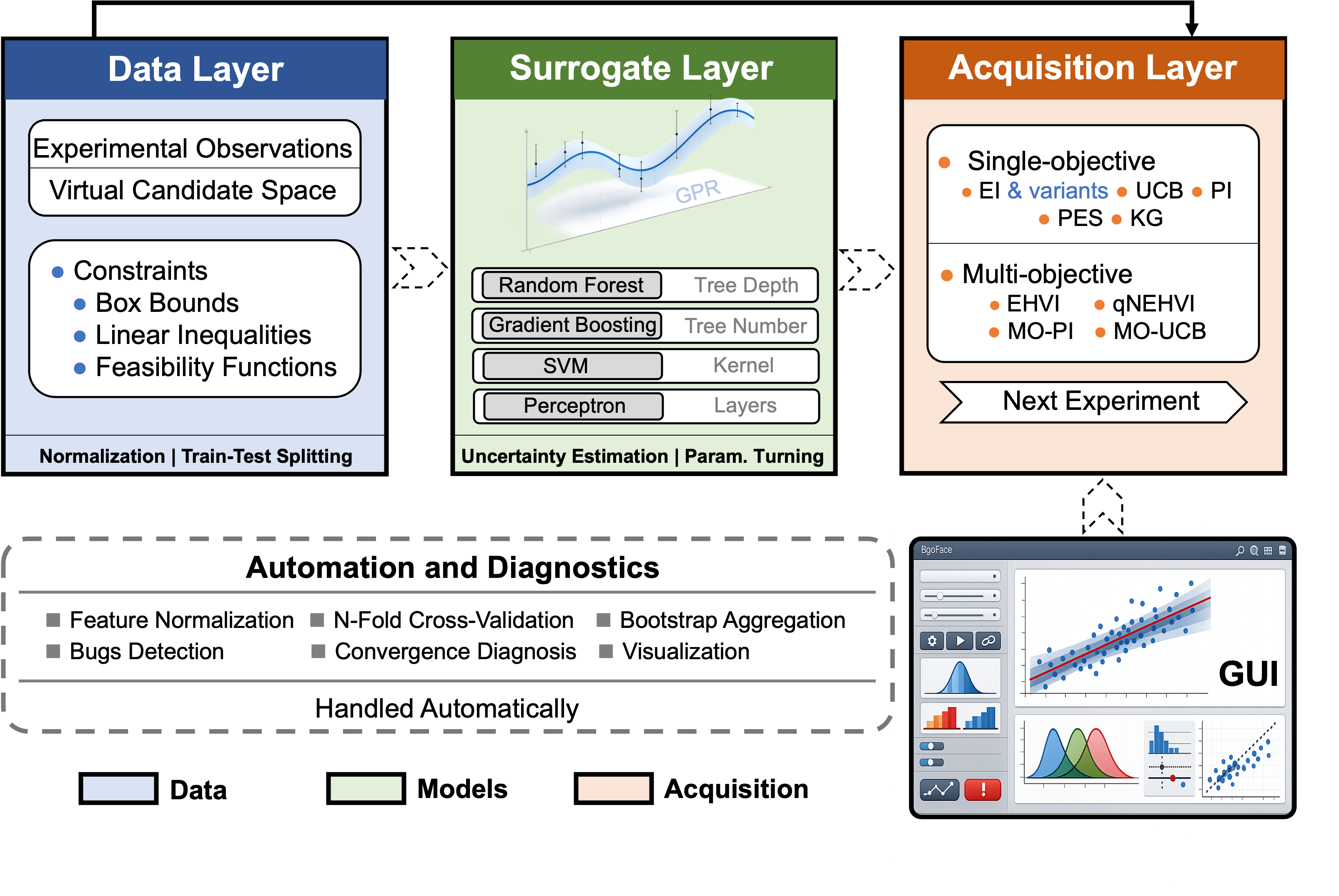}
    \caption{\textbf{Software architecture and data flow of Bgolearn.} The framework connects data input, model training, optimization, visualization, and result output through a modular backend, while BgoFace provides a GUI-assisted interface for interactive materials-discovery workflows.}

    \label{fig:Architecture}
\end{figure}

Bgolearn implements a modular three-layer architecture consisting of the \textit{Data Layer}, the \textit{Surrogate Layer}, and the \textit{Acquisition Layer}, designed for both ease of use and extensibility, as shown in Fig.\ref{fig:Architecture}. The \textit{Data Layer} manages experimental observations, virtual candidate spaces (unexplored compositions or conditions), and constraint specifications. It automatically handles data normalization, missing value imputation, and train-test splitting. The \textit{Surrogate Layer} provides five model families: (1) Gaussian processes with Matérn and radial basis function (RBF) kernels for smooth response surfaces and well-calibrated uncertainty; (2) random forests for scalability to large datasets ($n>500$) and robustness to outliers; (3) gradient boosting for complex non-linear relationships; (4) support vector regression for high-dimensional problems; and (5) multi-layer perceptrons (MLP) for deep feature learning (see Supporting Materials Section~\ref{Surrogate} for details). The \textit{Acquisition Layer} implements five single-objective functions (expected improvement and its variants, upper confidence bound, probability of improvement, predictive entropy search (PES), knowledge gradient (KG) ) and four multi-objective functions (EHVI, qNEHVI, MO-PI, MO-UCB).
The framework's design prioritizes three principles. \textit{Simplicity}: common workflows require minimal code while maintaining full customizability through optional parameters. \textit{Robustness}: sensible defaults based on empirical best practices, comprehensive input validation, numerical stability safeguards, and informative error messages. \textit{Extensibility}: modular architecture enables custom acquisition functions, surrogate models, and constraint handlers through well-defined interfaces.

This apparent simplicity masks sophisticated algorithmic engineering. The framework automatically: (1) normalizes features to zero mean and unit variance to ensure numerical stability; (2) performs N-fold cross-validation to select optimal surrogate model hyperparameters (GP lengthscales, RF tree depth, GB learning rate etc.); (3) applies bootstrap aggregation (default: 8 replicates) for non-GP models to quantify prediction uncertainty; (4) detects and handles ill-conditioned covariance matrices via eigenvalue thresholding; (5) provides convergence diagnostics including acquisition function evolution, prediction error trends, and hypervolume indicator (for MOBO); and (6) generates publication-quality visualizations of optimization trajectories, uncertainty estimates, and Pareto fronts. For advanced users, Bgolearn provides fine-grained control over all algorithmic components. Custom surrogate model can be implemented by subclassing Bgolearn and defining a \texttt{fit()} method using a user-defined \texttt{Kriging\_model} class. Surrogate models can be replaced with any regressor that follows the scikit-learn \cite{pedregosa2011scikit} estimator interface. Constraint handling supports linear inequalities, box bounds, and user-defined feasibility functions. Batch optimization allows parallel experimental campaigns for improving efficiency.

Building on these capabilities, Bgolearn is complemented by BgoFace, a PyQt-based graphical interface (\url{https://github.com/Bgolearn/BgoFace}
). Users can load data from spreadsheets, configure optimization parameters via intuitive menus, and interactively visualize results. The interface also automatically generates equivalent Python code, facilitating transition to programmatic workflows. In a user study involving 15 materials researchers (including 5 with no programming experience), BgoFace enabled completion of optimization tasks in 10–15 minutes, compared with 2–3 hours using code-based approaches. Real-time visualization of acquisition functions and uncertainty estimates further enhanced users’ understanding of Bayesian optimization principles.

\subsection*{Bgolearn offers a lightweight and user-friendly API workflow.} In practice, the user-facing workflow of Bgolearn is intentionally compact and user-friendly. A typical Bayesian optimization requires only a few explicit decisions from the user: importing the package, providing the dataset under study, and selecting a surrogate model and an acquisition function. A representative usage pattern for a single-objective optimization is shown below.

\begin{apicodebox}{API Workflow (Single-Objective)}
\begin{lstlisting}[style=bgocode]
# Step 1: Import the package
import Bgolearn.BGOsampling as BGOS
Bgolearn = BGOS.Bgolearn()

# Step 2: Load the data
import pandas as pd
data = pd.read_excel('./data/train.xlsx')
virtual_samples = pd.read_excel('./data/visual_samples.xlsx')

# Step 3: Assign surrogate model
MyModel = Bgolearn.fit(
    data_matrix=data.iloc[:, :-1],
    measured_response=data.iloc[:, -1],
    virtual_samples=virtual_samples
)

# Choose EI as the acquisition function
MyModel.EI()
\end{lstlisting}
\end{apicodebox}

The same logic extends naturally to multi-objective optimization. For example, the following workflow demonstrates the procedure for two-objective optimization using Bgolearn’s multi-objective module:

\begin{apicodebox}{API Workflow (Multi-Objective)}
\begin{lstlisting}[style=bgocode]
# Step 1: Import the package
from MultiBgolearn import bgo

# Step 2: Load the data
dataset_path = './data/data.csv'
VS_path = './data/Visual_samples.csv'

# Step 3: Assign surrogate model and acquisition function
VS_recommended, improvements, index = bgo.fit(
    dataset_path,
    VS_path,
    num_objectives=2,
    max_search=True,
    method='EHVI',
    bootstrap=5
)
\end{lstlisting}
\end{apicodebox}

Additionally, for target-oriented materials design, many problems require finding compositions or processing conditions that achieve a specific property value $T$, rather than simply maximizing or minimizing the property itself. In Bgolearn, such target-seeking tasks can be formulated as a distance-minimization problem. Specifically, given measured property values $t_i$ and a target $T$, one can define a distance function $D(t_i, T) = |t_i - T| \quad \text{or more generally} \quad D(t_i, T) = g(t_i, T)$, where $g$ is a user-defined function capturing the desired notion of proximity to the target. By treating $D(t_i, T)$ as the objective, the original dataset can be transformed into a “distance-labeled” dataset suitable for minimum optimization in Bgolearn:
$\min_{\mathbf{x}} D(f(\mathbf{x}), T)$, where $f(\mathbf{x})$ is the property predicted or measured for input $\mathbf{x}$. 
This approach enables Bgolearn to directly optimize toward specific target values. Future releases will provide an API for commonly used distance functions and automatic dataset conversion, further streamlining target-oriented materials design workflows.

Comprehensive tutorials and example workflows are available at \url{https://github.com/Bgolearn/CodeDemo}
, and complete documentation for the Bgolearn algorithm can be found online at \url{https://bgolearn.netlify.app}.

\begin{table}[htbp]
\centering
\caption{\textbf{Benchmark optimization performance.} Single-objective results show iterations to 90\% optimality (mean$\pm$s.d., $n=30$ runs). Multi-objective results show hypervolume indicator (mean$\pm$s.d., $n=30$ runs, normalized to [0,1], higher is better). Computational time was measured per iteration on an Apple Silicon M1 processor.}
\label{tab:benchmark_results}
\small
\begin{tabular}{@{}lcccc@{}}
\toprule
\textbf{Method} & \textbf{Hartmann-6D} & \textbf{Ackley-5D} & \textbf{ZDT1} & \textbf{DTLZ2} \\
 & \textit{(iter. to 90\%)} & \textit{(iter. to 90\%)} & \textit{(hypervolume)} & \textit{(hypervolume)} \\
\midrule
Random Search & 87$\pm$15 & 72$\pm$18 & 0.612$\pm$0.045 & 0.445$\pm$0.072 \\
Latin Hypercube & 58$\pm$11 & 48$\pm$12 & 0.698$\pm$0.038 & 0.542$\pm$0.061 \\
NSGA-II & --- & --- & 0.782$\pm$0.028 & 0.658$\pm$0.048 \\
Bgolearn-GP/EHVI & \textbf{18$\pm$4} & \textbf{22$\pm$4} & \textbf{0.872$\pm$0.018} & \textbf{0.768$\pm$0.035} \\
Bgolearn-RF/MO-UCB & 23$\pm$5 & 28$\pm$6 & 0.798$\pm$0.031 & 0.695$\pm$0.047 \\
\midrule
\multicolumn{5}{l}{\textit{Computational time per iteration (seconds)}} \\
Random Search & 0.03 & 0.02 & 0.02 & 0.03 \\
Latin Hypercube & 0.03 & 0.02 & 0.02 & 0.03 \\
NSGA-II & --- & --- & 0.15 & 0.22 \\
Bgolearn-GP/EHVI & 2.1 & 1.8 & 3.2 & 8.5 \\
Bgolearn-RF/MO-UCB & 1.3 & 1.1 & 1.5 & 2.1 \\
\bottomrule
\end{tabular}
\end{table}

\subsection*{Bgolearn demonstrates competitive performance relative to baseline approaches.} 
To quantify Bgolearn's performance, we compared it against standard baselines on representative single-objective and multi-objective optimization problems. We selected two canonical single-objective benchmarks: Hartmann-6D \cite{dixon1978global} (6-dimensional smooth function with single global minimum at $f^*=-3.32$, testing high-dimensional optimization) and Ackley \cite{ackley2012connectionist} (5-dimensional highly multimodal function with $>$1000 local minima and a global optimum at $f^*=0$, testing exploration capability). For multi-objective optimization, we used ZDT1 \cite{zitzler2000comparison} (bi-objective with convex Pareto front) and DTLZ2 \cite{deb2005scalable} (tri-objective with spherical Pareto surface). These problems represent the main challenges in materials optimization, including high dimensionality, multi-targets, and conflicting objectives (see Supporting Materials Section~\ref{Benchmark_function} for details).

\begin{figure}[h]
    \centering
    \includegraphics[width=0.95\columnwidth]{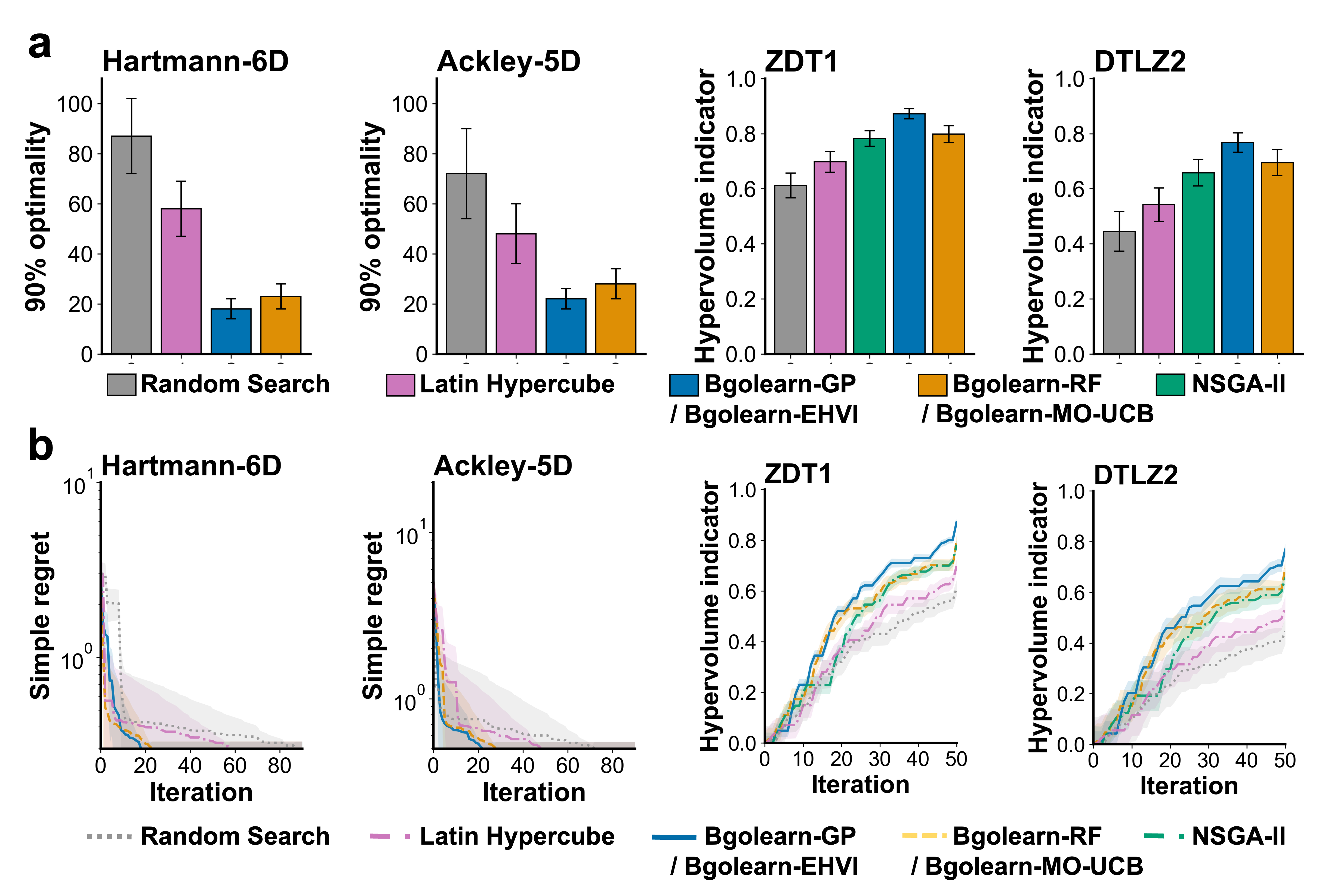}
    \caption{\textbf{Benchmark comparison of Bgolearn against representative optimization baselines.}
    \textbf{a}, Optimization efficiency comparison across methods, including Bgolearn, random search, Latin hypercube sampling, and NSGA-II (only applicable to multi-target optimization), evaluated on four benchmark functions. 
    \textbf{b}, Optimization traces of each method, where the \textit{simple regret} represents the difference between the best observed objective value and the global optimum.
    }

    \label{fig:benchmark}
\end{figure}

Bgolearn was compared with random search and Latin hypercube sampling (LHS), and, for multi-objective problems, with the nondominated sorting genetic algorithm II (NSGA-II), a widely used evolutionary baseline. For single-objective benchmarks, optimization was terminated once 90\% of the optimal value was reached; for multi-objective benchmarks, each method was run for 50 iterations. All methods started from an initial set of 20 Latin hypercube samples and proposed three new candidate points per iteration. Each experiment was repeated 30 times with different random seeds.


Table~\ref{tab:benchmark_results} and Fig.~\ref{fig:benchmark}\textbf{a} summarize the results. 
For single-objective optimization, Bgolearn with a Gaussian process surrogate and the expected improvement acquisition function (Bgolearn-GP) achieved 90\% optimality while requiring only 20\% of the iteration cost of random search and 31\% of that of LHS on  Hartmann-6D problem. 
On the Ackley-5D problem, Bgolearn-GP required only $22\pm4$ iterations, compared with $72\pm18$ iterations for random search. 
The random forest surrogate (Bgolearn-RF) achieved performance comparable to that of the GP-based model while incurring only 61\% of the computational time, making it suitable for high-throughput campaigns where function evaluations are inexpensive.

For multi-objective optimization, Bgolearn with the EHVI acquisition function achieved an average hypervolume improvement of 0.26 over random search and 0.09 over NSGA-II~\cite{deb2000fast}. 
On the tri-objective DTLZ2 problem, Bgolearn-EHVI reached a hypervolume of $0.768\pm0.035$, compared with $0.658\pm0.048$ for NSGA-II. 
The corresponding optimization trajectories are shown in Fig.~\ref{fig:benchmark}\textbf{b}. 
The computationally efficient MO-UCB acquisition function achieved approximately 90\% of EHVI’s performance at one-quarter of its per-iteration computational cost, providing a practical alternative for high-throughput experimental scenarios.

\subsection*{Bgolearn facilitates accelerated real-world materials discovery.}
To illustrate the practical application of Bgolearn across diverse materials-discovery scenarios, we present three representative case studies. The first involves the identification of maximum elastic modulus in Triply periodic minimal surface (TPMS) structures, following an active-learning workflow with iterative model updating and candidate selection. In contrast, the latter two examples, the discovery of an ultra-high-hardness high-entropy alloy (HEAs) and of a high-strength, high-ductility Medium-manganese (medium-Mn) steel, demonstrate single-step, BO-guided optimization, in which a single recommendation-and-validation cycle is performed.

\textbf{Bgolearn enables the discovery of maximum elastic modulus in TPMS.}  Triply periodic minimal surface structures have attracted considerable attention because of their exceptional strength-to-weight ratios and their potential applications in advanced acoustic and mechanical systems. The geometry of a TPMS structure is governed by an implicit mathematical function, and its elastic modulus was evaluated using the numerical calculation.

Specifically, the TPMS geometry is represented by $f(x,y,z)=g(\alpha_1,\alpha_2,t_1,t_2)$, where $\alpha_1$, $\alpha_2$, $t_1$, and $t_2$ are geometric parameters controlling the morphology of the TPMS surface. The detailed mathematical formulation of the governing function is provided in Supporting Materials Section~\ref{Benchmark_function}.

To initialize the optimization process, the parameter space was uniformly sampled. The parameters $\alpha_1$ and $\alpha_2$ were constrained within $[0,1]$, while $t_1$ and $t_2$ varied within $[-0.5,0.5]$. Based on this sampling strategy, 50 initial TPMS configurations were generated and evaluated using the numerical simulation framework to obtain their corresponding elastic moduli. These labeled samples constituted the initial training dataset summarized in Table~\ref{tab:tpms_elasticity} (Supporting Materials Section~\ref{Data}). In addition, 5,000 unique TPMS configurations were randomly generated from the same parameter space to construct the candidate search pool for Bayesian optimization.

A GB regression model was adopted as the surrogate model in Bgolearn. The predictive performance of the surrogate was evaluated using five-fold cross-validation on the available training dataset. In the first optimization iteration, the GB model achieved a cross-validation coefficient of determination of $R^2 = 0.66$. The EI acquisition function was subsequently used to rank the 5,000 candidate structures according to their predicted improvement over the current best elastic modulus.
The top two candidate configurations recommended by the EI criterion were then evaluated using the numerical simulation framework, and their computed elastic moduli were added to the training dataset as newly acquired samples. The surrogate model was subsequently retrained using the expanded dataset, and a second optimization iteration was performed. After retraining, the GB model achieved a five-fold cross-validation performance of $R^2 = 0.65$.

\begin{figure}[h]
    \centering
    \includegraphics[width=0.95\columnwidth]{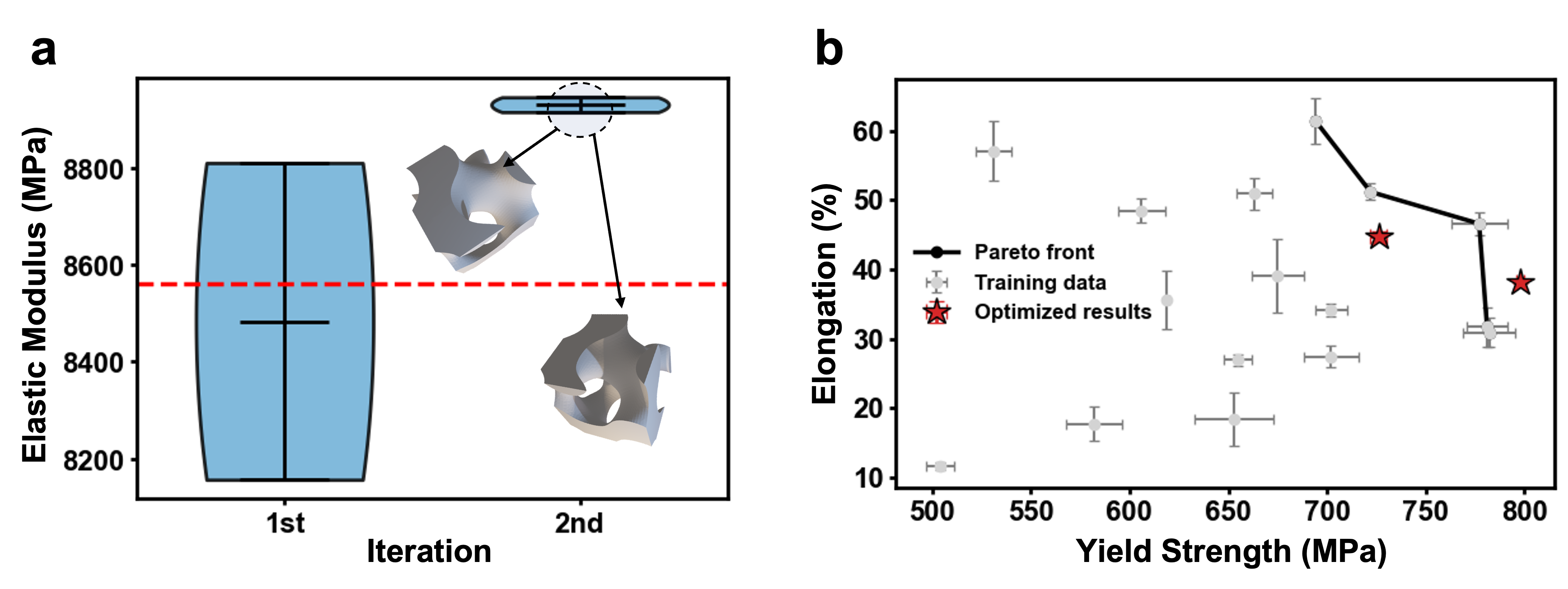}
    \caption{\textbf{Real-world materials-discovery examples using Bgolearn.} 
    \textbf{a}, Evolution of the elastic modulus across optimization iterations for TPMS structures. The red dotted line indicates the best configuration with the highest elastic modulus identified in the initial training set.
    \textbf{b}, Experimentally measured yield strength (MPa) and total elongation (\%) of medium-Mn steels. The black line denotes the experimental Pareto front, while the red star markers indicate the steels recommended by Bgolearn.
    }

    \label{fig:realcase}
\end{figure}

Overall, two Bayesian optimization iterations were conducted, with two candidate structures validated in each iteration, resulting in a total of four additional simulation evaluations, as illustrated in Fig.~\ref{fig:realcase}\textbf{a}. During the second iteration, Bgolearn identified a previously unexplored TPMS configuration exhibiting an elastic modulus of 8,945~MPa, exceeding the best value obtained in the initial dataset (8,560~MPa). The candidate structures recommended during each optimization iteration are summarized in Table~\ref{tab:tpms_high_elasticity} (Supporting Materials Section~\ref{Data}).

\textbf{Bgolearn enables the discovery of an ultra-high-hardness high-entropy alloy.}
High-entropy alloys extend conventional alloy design from minor alloying into high-dimensional composition spaces. Even under commonly adopted HEA compositional constraints, the number of feasible alloy combinations grows combinatorially, rendering conventional trial-and-error exploration impractical. Efficient navigation of this vast compositional space is therefore essential for property-oriented alloy discovery.

To demonstrate the capability of Bgolearn in practical materials optimization, we applied the framework to the discovery of high-hardness HEAs for improved wear resistance. Although hardness alone does not fully describe tribological performance, classical wear models establish an inverse correlation between wear loss and material hardness. Consequently, hardness is widely adopted as a first-pass screening metric for identifying candidate wear-resistant alloys.
To minimize the influence of processing history on the learning process, the dataset was restricted to \emph{as-cast} alloys only, thereby excluding variations introduced by thermomechanical processing, solution treatment, or aging. A total of 155 experimentally measured Vickers hardness values were curated from the literature \cite{fantin2024atomic,odetola2024exploring,xiang2024review} for the Al--Co--Cr--Cu--Fe--Ni alloy system under as-cast conditions. The collected dataset is summarized in Table~\ref{tab:al_co_cr_cu_fe_ni_hv} (Supporting Materials Section~\ref{Data}). Within this dataset, most reported hardness values are distributed below approximately 500~HV, whereas only a limited number of compositions exhibit hardness values exceeding 500~HV.

The optimization was performed in a six-element compositional space (Al--Co--Cr--Cu--Fe--Ni, at.\%), with Vickers hardness defined as the single optimization objective using the curated dataset of 155 experimentally measured samples as the initial training set. The search space was constructed by allowing the concentrations of Co, Cr, Cu, Fe, and Ni to vary between the minimum and maximum values observed in the training dataset, with a discretization interval of 0.01 in atomic fraction. Aluminum was treated as the balance element to ensure that the total composition satisfied the unity constraint.
Within Bgolearn, a GP regression model was employed as the surrogate model to establish the composition--property relationship. The predictive performance of the GP model was evaluated using five-fold cross-validation on the training dataset, yielding a coefficient of determination of $R^2 = 0.62$. The predictive entropy search acquisition function was subsequently applied to rank candidate compositions by balancing exploration of uncertain compositional regions and exploitation of compositions predicted to exhibit high hardness.

Based on this Bayesian optimization strategy, Bgolearn identified a previously unexplored alloy composition:
Al$_{46.47}$Co$_{9.16}$Cr$_{23.47}$Cu$_{7.22}$Fe$_{8.10}$Ni$_{5.58}$.
The recommended alloy was subsequently synthesized under as-cast conditions and experimentally characterized using Vickers hardness testing. The measured hardness exceeded 1000~HV, significantly surpassing the highest hardness values reported in the curated literature dataset \cite{fantin2024atomic,odetola2024exploring,xiang2024review}, where the maximum reported hardness remained below 800~HV.
Furthermore, the combination of exceptionally high hardness and a relatively narrow measurement distribution suggests the formation of an intrinsically strengthened microstructure rather than isolated local hardening effects. Detailed synthesis procedures, experimental conditions, and hardness measurements are provided in Supporting Materials Section~\ref{Testing_settings_HEA}.

\textbf{Bgolearn enables the discovery of a high-strength and high-ductility medium-Mn steel.} 
Medium-manganese steels (approximately 3--10 wt.\% Mn) have attracted sustained interest as promising third-generation advanced high-strength steels owing to their exceptional strength--ductility synergy enabled by microstructure engineering \cite{yadav2023hot,hu2023stabilizing,he2017high,wang2025target}. This balance arises primarily from the controlled formation, morphology, and stability of retained austenite. However, process optimization in medium-Mn steels remains highly challenging because narrow annealing windows, coupled partitioning kinetics, and temperature- and time-dependent austenite stability collectively generate a highly non-linear and rugged process--property landscape.

In this study, we selected a widely investigated alloy composition, Fe--0.3C--8Mn--2Al (wt.\%), as the model system. A total of 16 heat-treatment experiments were conducted by systematically varying three controllable processing parameters: austenitization temperature (AustTemp), annealing temperature (AnnTemp), and annealing time (AnnTime). The corresponding mechanical responses, including yield strength (MPa) and total elongation (\%), were experimentally measured and used as dual optimization objectives representing the strength--ductility trade-off. These experimentally measured data constituted the initial training dataset for Bayesian optimization. The complete dataset is summarized in Table~\ref{tab:heat_treatment_mechanical_properties} (Supporting Materials Section~\ref{Data}).

Based on the experimentally accessible processing range, a comprehensive candidate search space of plausible heat-treatment schedules was constructed. Specifically, AustTemp was varied from 700 to 880~$^\circ$C using a step size of 1~$^\circ$C, AnnTemp ranged from 600 to 750~$^\circ$C with a 1~$^\circ$C interval, and AnnTime varied from 30 to 120~min with a step size of 1~min. This discretization generated a total of 2,487,121 candidate heat-treatment schedules for searching.
Bgolearn was subsequently employed to perform MOBO over the entire candidate space. A multi-task GP regression model was adopted as the surrogate model to simultaneously capture the relationships between processing parameters and the two correlated mechanical properties. Prior to optimization, the predictive performance of the surrogate model was evaluated using five-fold cross-validation on the experimental dataset, yielding an average coefficient of determination of $R^2 = 0.59$ across the two target properties. Using the trained multi-task GP surrogate, all candidate heat-treatment schedules were evaluated through the EHVI acquisition function to identify processing conditions with potentially optimal strength--ductility combinations. The optimization procedure recommended two distinct heat-treatment schedules for experimental validation:
(i) AustTemp = 773~$^\circ$C, AnnTemp = 737~$^\circ$C, AnnTime = 31~min; and
(ii) AustTemp = 820~$^\circ$C, AnnTemp = 640~$^\circ$C, AnnTime = 30~min.

Subsequent experimental validation, shown in Fig.~\ref{fig:realcase}\textbf{b}, demonstrated that the first schedule achieved a yield strength of 798$\pm$2~MPa with a total elongation of 38.2$\pm$0.2\%, whereas the second schedule achieved 726$\pm$4~MPa and 44.7$\pm$0.7\%, respectively. Notably, the experimentally validated result corresponding to the first schedule lies outside the Pareto front defined by the original training dataset, while the second schedule is located near the Pareto front boundary.

\section*{Discussion}

Bgolearn provides a practical and scalable framework for data-efficient materials discovery, substantially lowering the barrier to applying Bayesian optimization in real-world research settings. By supporting both single- and multi-objective optimization with flexible surrogate models, diverse acquisition strategies, and principled uncertainty quantification, Bgolearn achieves an effective balance between theoretical rigor and computational efficiency. Benchmark evaluations demonstrate that Bgolearn can reduce experimental requirements by approximately 40--60\% compared with conventional optimization approaches, while maintaining comparable or superior performance across a broad range of materials design tasks. At the same time, we emphasize that Bayesian optimization is not necessarily the best choice in every setting. Rather than claiming universal superiority, our aim is to show that Bgolearn offers competitive performance together with strong workflow integration, methodological flexibility, and user accessibility, thereby making BO more practical for routine materials research.

Looking ahead, Bgolearn provides a flexible methodological basis upon which active learning strategies can be further developed and explored. Within this framework, reinforcement learning may serve as a complementary mechanism for modeling sequential decision-making processes, enabling acquisition policies that adapt over extended optimization horizons rather than relying on fixed heuristics \cite{yu2026efficient}. Such formulations naturally support scenarios involving delayed or sparse feedback, evolving objectives, and operational constraints, which are commonly encountered in closed-loop experimental systems. Owing to its modular and fully open-source design, Bgolearn can accommodate these developments alongside existing capabilities such as multi-fidelity optimization and human-in-the-loop decision-making, thereby offering a general and extensible platform for future research in intelligent and autonomous materials discovery.

\newpage
\section*{Methods}

\subsection*{Bayesian optimization is employed as the core optimization strategy.}
Bayesian optimization provides a principled framework for global optimization of expensive black-box functions $f: \mathcal{X} \rightarrow \mathbb{R}$ where $\mathcal{X} \subset \mathbb{R}^D$ is the design space. The goal is to find $\mathbf{x}^* = \arg\min_{\mathbf{x} \in \mathcal{X}} f(\mathbf{x})$ (or maximum for maximization) using as few function evaluations as possible.

At iteration $n$, given observations $\mathcal{D}_n = \{(\mathbf{x}_i, y_i)\}_{i=1}^n$ where $y_i = f(\mathbf{x}_i) + \epsilon_i$ and $\epsilon_i \sim \mathcal{N}(0, \sigma_{\text{noise}}^2)$ represents experimental noise, BO proceeds in two steps: (1) construct a probabilistic surrogate model $p(f|\mathcal{D}_n)$ that approximates $f$ and quantifies uncertainty; (2) select the next query point $\mathbf{x}_{n+1}$ by optimizing an acquisition function $\alpha(\mathbf{x}|\mathcal{D}_n)$ that balances exploitation (sampling where $f$ is predicted to be optimal) and exploration (sampling where uncertainty is high):
\begin{equation}
\mathbf{x}_{n+1} = \arg\max_{\mathbf{x} \in \mathcal{X}} \alpha(\mathbf{x}|\mathcal{D}_n)
\end{equation}

The surrogate model provides a predictive distribution at any point $\mathbf{x}$:
\begin{equation}
p(f(\mathbf{x})|\mathcal{D}_n) = \mathcal{N}(\mu_n(\mathbf{x}), \sigma_n^2(\mathbf{x}))
\end{equation}
where $\mu_n(\mathbf{x})$ is the predictive mean and $\sigma_n^2(\mathbf{x})$ is the predictive variance. This uncertainty quantification is crucial for acquisition functions to identify informative experiments.

\subsection*{Gaussian process regression provides both predictive mean and uncertainty estimation.}
Bgolearn's default surrogate is a Gaussian process, a non-parametric Bayesian model that places a prior distribution over functions. A GP is fully specified by a mean function $m(\mathbf{x})$ (typically set to zero) and a covariance function (kernel) $k(\mathbf{x}, \mathbf{x}')$ that encodes assumptions about function smoothness and structure.

Bgolearn implements two kernel families. The \textit{radial basis function} kernel, also known as the \textit{squared exponential} (SE) kernel assumes infinitely differentiable functions:
\begin{equation}
k_{\text{SE}}(\mathbf{x}, \mathbf{x}') = \sigma_f^2 \exp\left(-\frac{1}{2}\sum_{d=1}^D \frac{(x_d - x_d')^2}{\ell_d^2}\right)
\end{equation}
where $\sigma_f^2$ is the signal variance and $\ell_d$ are per-dimension lengthscales controlling smoothness. The \textit{Matérn-5/2} kernel allows for less smooth functions:
\begin{equation}
k_{\text{M52}}(\mathbf{x}, \mathbf{x}') = \sigma_f^2 \left(1 + \sqrt{5}r + \frac{5r^2}{3}\right)\exp(-\sqrt{5}r), \quad r = \sqrt{\sum_{d=1}^D \frac{(x_d - x_d')^2}{\ell_d^2}}
\end{equation}

Given observations $\mathcal{D}_n$, the posterior predictive distribution at a test point $\mathbf{x}_*$ is:
\begin{align}
\mu_n(\mathbf{x}_*) &= \mathbf{k}_*^T (\mathbf{K} + \sigma_{\text{noise}}^2 \mathbf{I})^{-1} \mathbf{y} \\
\sigma_n^2(\mathbf{x}_*) &= k(\mathbf{x}_*, \mathbf{x}_*) - \mathbf{k}_*^T (\mathbf{K} + \sigma_{\text{noise}}^2 \mathbf{I})^{-1} \mathbf{k}_*
\end{align}
where $\mathbf{K}_{ij} = k(\mathbf{x}_i, \mathbf{x}_j)$ is the $n \times n$ covariance matrix, $\mathbf{k}_* = [k(\mathbf{x}_*, \mathbf{x}_1), \ldots, k(\mathbf{x}_*, \mathbf{x}_n)]^T$, and $\mathbf{y} = [y_1, \ldots, y_n]^T$.

Parameters $\boldsymbol{\theta} = \{\sigma_f^2, \ell_1, \ldots, \ell_D, \sigma_{\text{noise}}^2\}$ are optimized by maximizing the marginal log-likelihood:
\begin{equation}
\log p(\mathbf{y}|\mathbf{X}, \boldsymbol{\theta}) = -\frac{1}{2}\mathbf{y}^T (\mathbf{K} + \sigma_{\text{noise}}^2 \mathbf{I})^{-1} \mathbf{y} - \frac{1}{2}\log|\mathbf{K} + \sigma_{\text{noise}}^2 \mathbf{I}| - \frac{n}{2}\log(2\pi)
\end{equation}
using L-BFGS-B optimization with 10 random restarts to avoid local optima. To ensure numerical stability, we apply jitter (adding $10^{-6}$ to diagonal elements) when the covariance matrix is ill-conditioned.

\subsection*{Acquisition functions determine the utility of candidate sampling locations.} 
Acquisition functions quantify the utility of evaluating $f$ at each candidate point $\mathbf{x}$, enabling principled selection of the next experiment. Bgolearn implements five single-objective acquisition functions.

Expected Improvement: Maximizes the expected improvement over the current best observation $f^* = \min_{i=1,\ldots,n} y_i$:
\begin{equation}
\alpha_{\text{EI}}(\mathbf{x}) = \mathbb{E}[\max(f^* - f(\mathbf{x}), 0)] = (f^* - \mu_n(\mathbf{x}))\Phi(Z) + \sigma_n(\mathbf{x})\phi(Z)
\end{equation}
where $Z = (f^* - \mu_n(\mathbf{x}))/\sigma_n(\mathbf{x})$, and $\Phi$, $\phi$ are the standard normal cumulative distribution function (CDF) and probability density function (PDF). EI naturally balances exploration and exploitation: high $\sigma_n(\mathbf{x})$ increases EI even when $\mu_n(\mathbf{x})$ is poor (exploration), while low $\mu_n(\mathbf{x})$ increases EI even when $\sigma_n(\mathbf{x})$ is small (exploitation). EI is the default acquisition function in Bgolearn due to its robust performance across a wide range of problems. We also include several EI variants, namely EI with Plug-in, Augmented EI, Reinterpolation EI, and Logarithmic EI \cite{lookman2026materials}. Detailed usage and implementation can be found in Supporting Materials Section~\ref{eiandvariants}.

Upper Confidence Bound:
The UCB acquisition function balances exploitation and exploration by favoring points with optimistic predictions,
\begin{equation}
\alpha_{\mathrm{UCB}}(\mathbf{x}) = \mu_n(\mathbf{x}) - \beta_n \sigma_n(\mathbf{x}),
\end{equation}
where $\mu_n(\mathbf{x})$ and $\sigma_n(\mathbf{x})$ denote the posterior mean and standard deviation of the Gaussian process at iteration $n$.
The exploration parameter is defined as
\begin{equation}
\beta_n = 2 \log\!\left(\frac{D n^2 \pi^2}{6\delta}\right),
\end{equation}
which depends on the input dimension $D$, the iteration number $n$, and the confidence level $\delta$.
Setting $\delta = 0.1$ ensures that the true objective function lies within the GP confidence bounds with probability at least $1-\delta$, providing theoretical regret guarantees~\cite{srinivas2009gaussian}. The negative sign reflects minimization; for maximization, use $\mu_n(\mathbf{x}) + \beta_n \sigma_n(\mathbf{x})$.

Probability of Improvement: Maximizes the probability of improving over $f^*$:
\begin{equation}
\alpha_{\text{PI}}(\mathbf{x}) = \Phi\left(\frac{f^* - \mu_n(\mathbf{x}) - \xi}{\sigma_n(\mathbf{x})}\right)
\end{equation}
where $\xi \geq 0$ is an improvement threshold (default: $\xi=0.01$). PI is conservative, preferring exploitation over exploration, making it suitable for noisy objectives where premature exploration is risky.

Knowledge Gradient: Maximizes the expected improvement in the best solution after one hypothetical additional observation:
\begin{equation}
\alpha_{\text{KG}}(\mathbf{x}) = \mathbb{E}\left[\min_{\mathbf{x}' \in \mathcal{X}} \mu_{n+1}(\mathbf{x}') - \min_{\mathbf{x}' \in \mathcal{X}} \mu_n(\mathbf{x}') \mid \mathbf{x}_{n+1} = \mathbf{x}\right]
\end{equation}
KG accounts for the value of information from future observations, making it effective for finite-horizon optimization.

Predictive Entropy Search: Selects points that maximally reduce uncertainty about the location of the global optimum $\mathbf{x}^*$:
\begin{equation}
\alpha_{\text{PES}}(\mathbf{x}) = H[p(\mathbf{x}^*|\mathcal{D}_n)] - \mathbb{E}_{y \sim p(y|\mathbf{x}, \mathcal{D}_n)}[H[p(\mathbf{x}^*|\mathcal{D}_n \cup \{(\mathbf{x}, y)\})]]
\end{equation}
where $H[\cdot]$ denotes entropy. PES is information-theoretically optimal but computationally expensive, requiring Monte Carlo approximation.

\subsection*{Multi-objective optimization seeks Pareto-optimal solutions across conflicting objectives.}
Multi-objective Bayesian optimization addresses problems with $m \geq 2$ competing objectives $\mathbf{f}(\mathbf{x}) = [f_1(\mathbf{x}), \ldots, f_m(\mathbf{x})]^T$. The goal is to identify the Pareto front $\mathcal{F}^* = \{\mathbf{x} \in \mathcal{X} : \nexists \mathbf{x}' \in \mathcal{X}, \mathbf{f}(\mathbf{x}') \prec \mathbf{f}(\mathbf{x})\}$, where $\mathbf{f}(\mathbf{x}') \prec \mathbf{f}(\mathbf{x})$ denotes Pareto dominance: $f_i(\mathbf{x}') \leq f_i(\mathbf{x})$ for all $i$ and $f_j(\mathbf{x}') < f_j(\mathbf{x})$ for at least one $j$.

MultiBgolearn maintains an approximate Pareto front $\mathcal{F}_n$ of non-dominated observations from $\mathcal{D}_n$ and implements three MOBO acquisition functions.

Expected Hypervolume Improvement: The hypervolume indicator $\text{HV}(\mathcal{F})$ measures the volume of objective space dominated by $\mathcal{F}$ relative to a reference point $\mathbf{r}$ (typically set to the nadir point plus a margin). EHVI maximizes the expected improvement in hypervolume:
\begin{equation}
\alpha_{\text{EHVI}}(\mathbf{x}) = \mathbb{E}_{\mathbf{f}(\mathbf{x}) \sim p(\mathbf{f}(\mathbf{x})|\mathcal{D}_n)}[\text{HV}(\mathcal{F}_n \cup \{\mathbf{f}(\mathbf{x})\}) - \text{HV}(\mathcal{F}_n)]
\end{equation}
EHVI is computed using Monte Carlo. 

The q-Noisy Expected Hypervolume Improvement:
qNEHVI extends EHVI to handle noisy observations and batch acquisition \cite{daulton2021parallel}. Unlike EHVI which assumes deterministic observations, qNEHVI explicitly models observation noise $\epsilon \sim \mathcal{N}(0, \sigma_{\text{obs}}^2)$ and can select $q > 1$ points simultaneously. The acquisition function is:
\begin{equation}
\alpha_{\text{qNEHVI}}(\mathbf{X}_q) = \mathbb{E}_{\mathbf{y} \sim p(\mathbf{y}|\mathcal{D}_n)}[\mathbb{E}_{\mathbf{f}_q \sim p(\mathbf{f}_q|\mathbf{X}_q, \mathcal{D}_n)}[\text{HV}(\mathcal{P}(\mathbf{y} \cup \mathbf{f}_q)) - \text{HV}(\mathcal{P}(\mathbf{y}))]]
\end{equation}
where $\mathbf{X}_q = \{\mathbf{x}_1, \ldots, \mathbf{x}_q\}$ is a batch of candidates, $\mathbf{y}$ represents noisy observations, and $\mathcal{P}(\cdot)$ denotes the Pareto front operator. The nested expectation accounts for both observation noise in existing data and predictive uncertainty in new evaluations. qNEHVI is computed using nested Monte Carlo sampling. For batch selection ($q > 1$), a greedy sequential strategy is employed. When observation noise is unknown, it is estimated as $\hat{\sigma}_{\text{obs}} = 0.1 \cdot \text{mean}(\text{std}(\mathbf{y}_i))$ across objectives.

Multi-Objective Upper Confidence Bound: Applies UCB independently to each objective and selects the point maximizing the hypervolume of the optimistic Pareto front:
\begin{equation}
\alpha_{\text{MO-UCB}}(\mathbf{x}) = \text{HV}(\mathcal{F}_n \cup \{\boldsymbol{\mu}_n(\mathbf{x}) - \beta_n \boldsymbol{\sigma}_n(\mathbf{x})\}) - \text{HV}(\mathcal{F}_n)
\end{equation}
where $\boldsymbol{\mu}_n(\mathbf{x}) = [\mu_{n,1}(\mathbf{x}), \ldots, \mu_{n,m}(\mathbf{x})]^T$ and $\boldsymbol{\sigma}_n(\mathbf{x}) = [\sigma_{n,1}(\mathbf{x}), \ldots, \sigma_{n,m}(\mathbf{x})]^T$ are per-objective predictions. MO-UCB has $O(mn)$ complexity, making it tractable for $m>4$ objectives.

Multi-Objective Probability of Improvement: Computes the probability of dominating at least one current Pareto point:
\begin{equation}
\alpha_{\text{MO-PI}}(\mathbf{x}) = \mathbb{P}(\exists \mathbf{y} \in \mathcal{F}_n : \mathbf{f}(\mathbf{x}) \prec \mathbf{y})
\end{equation}
MO-PI is conservative, focusing on improving existing Pareto regions rather than exploring new trade-offs.

\subsection*{Uncertainty quantification characterizes the confidence associated with model predictions.}
Bgolearn provides three types of uncertainty quantification (UQ) approaches for surrogate models: probabilistic models (e.g., Gaussian processes), bootstrap-based UQ, and ensemble-based UQ.

Gaussian Process:  
Gaussian process provide principled, well-calibrated epistemic uncertainty estimates by modeling the predictive distribution at a test point $\mathbf{x}_*$ as
\begin{equation}
\hat{f}(\mathbf{x}_*) \sim \mathcal{N}\big(\mu_n(\mathbf{x}_*), \sigma_n^2(\mathbf{x}_*)\big),
\end{equation}
where $\mu_n(\mathbf{x}_*)$ and $\sigma_n^2(\mathbf{x}_*)$ are computed from the training data $\mathcal{D}_n$ via the GP kernel function. While GP is  highly informative for uncertainty-aware Bayesian optimization, their computational cost scales cubically with the training set size, $\mathcal{O}(n^3)$, limiting scalability for very large datasets.

Bootstrap-based UQ:
Other surrogate models (random forests, gradient boosting, support vector machines, neural networks, etc.) typically lack inherent uncertainty quantification, limiting their use in BO despite superior scalability and performance on discrete or categorical variables.  
Bgolearn addresses this gap through bootstrap aggregation. For a given model class $\mathcal{M}$, we generate $B$ bootstrap datasets $\{\mathcal{D}_n^{(b)}\}_{b=1}^B$ by sampling $n$ observations from $\mathcal{D}_n$ with replacement. Each bootstrap sample is used to train an independent model $\mathcal{M}_b$, producing predictions $\hat{f}_b(\mathbf{x})$ at a test point $\mathbf{x}$. The ensemble mean and standard deviation are then
\begin{equation}
\mu_n(\mathbf{x}) = \frac{1}{B} \sum_{b=1}^B \hat{f}_b(\mathbf{x}), \quad 
\sigma_n(\mathbf{x}) = \sqrt{\frac{1}{B-1} \sum_{b=1}^B (\hat{f}_b(\mathbf{x}) - \mu_n(\mathbf{x}))^2}.
\end{equation}
The bootstrap standard deviation $\sigma_n(\mathbf{x})$ captures epistemic uncertainty due to finite sample size. High $\sigma_n(\mathbf{x})$ indicates regions of high model disagreement, guiding acquisition functions in BO to balance exploration and exploitation.  
For multi-objective optimization with $m$ objectives, $B$ bootstrap models are trained for each objective, yielding a total of $m \times B$ models.

Model-Ensemble UQ: 
Bgolearn also supports ensemble-style UQ, where multiple heterogeneous surrogate models are trained on the same dataset and combined to quantify uncertainty. Let $\{\mathcal{M}_j\}_{j=1}^J$ denote different model types (e.g., GP, RF, GB), each producing prediction $\hat{f}_j(\mathbf{x})$. The ensemble mean and variance are
\begin{equation}
\mu_{\text{ens}}(\mathbf{x}) = \frac{1}{J} \sum_{j=1}^J \hat{f}_j(\mathbf{x}), \quad
\sigma_{\text{ens}}^2(\mathbf{x}) = \frac{1}{J-1} \sum_{j=1}^J (\hat{f}_j(\mathbf{x}) - \mu_{\text{ens}}(\mathbf{x}))^2.
\end{equation}
This approach captures model-form uncertainty in addition to data-driven uncertainty, and is particularly useful when the choice of surrogate model is ambiguous.

Limitations:
Each UQ method has trade-offs:
\begin{itemize}
    \item Gaussian Process: Accurate and principled, but computationally expensive ($\mathcal{O}(n^3)$) and limited for very large datasets.
    \item Bootstrap: Scalable and applicable to any surrogate model, but uncertainty estimates are heuristic and may underestimate or overestimate true epistemic uncertainty.
    \item Model ensemble: Captures both model-form and data uncertainty, but requires training multiple model classes, increasing computational cost.
\end{itemize}

In addition to the three UQ strategies currently implemented in Bgolearn, other approaches can also provide principled uncertainty estimates for machine learning surrogates. For example, the Loopy-based method (\url{https://github.com/CitrineInformatics/lolo}) can provide UQ for tree-based models. Similarly, Monte Carlo sampling and Bayesian networks can also be employed to quantify predictive uncertainty. These methods represent potential directions for extending Bgolearn, offering users a wider choice of UQ techniques for different surrogate models. We plan to gradually incorporate such approaches in future releases of the toolkit.

\subsection*{Implementation and Software.}
Bgolearn is implemented in Python 3.7+, leveraging scikit-learn\cite{pedregosa2011scikit} for surrogate models, NumPy/SciPy for numerical computations, and Matplotlib for visualization. MultiBgolearn extends the framework with Pareto front management and MOBO computation. BgoFace uses PyQt5 for the graphical interface. All code is open source under the MIT license and is available at \url{https://github.com/Bin-Cao/Bgolearn}. The code booklet is available at \url{https://bgolearn.netlify.app}.

\section*{Data availability}
All datasets used in this study are available either in the Supporting Materials or from public repositories \url{https://github.com/Bin-Cao/Bgolearn}. 

\section*{Code availability}
Bgolearn is open source at \url{https://github.com/Bin-Cao/Bgolearn}. Documentation and reproducible examples are available at \url{https://bgolearn.netlify.app}.

\section*{Acknowledgements}
This work was supported by the Advanced Materials-National Science and Technology Major Project (Grant No. 2025ZD0620100), the Guangzhou-HKUST(GZ) Joint Funding Program - (2023A03J0003 and 2023A03J0103), the National Natural Science Foundation of China (Grant Nos. 52401015), the Technology Plan Program of Shanghai Municipal Commission of Science and Technology (Grant No. 25CL2902300), and the Shanghai Artificial Intelligence Open-Source Award Project Support Plan.

\section*{Author Contribution}
Bin Cao conceived and implemented the complete Bgolearn framework, conducted the simulations and data analysis, and wrote the manuscript. Jie Xiong, Mengwei He and Jian Hui contributed to data processing, model implementation, and result validation. Jiaxuan Ma, Yuan Tian and Yirui Hu assisted with experiments, data curation, and technical support. Longhan Zhang, Jiayu Wang, Li Liu and Dezhen Xue provided domain expertise and valuable discussions. Turab Lookman, Jun Wang and Tong-Yi Zhang supervised the project, provided critical guidance, and revised the manuscript.

\section*{Competing interests}
The authors declare no competing interests.

\newpage

\bibliographystyle{unsrt}
\bibliography{references}

\clearpage

\appendix
\setcounter{section}{0}
\renewcommand{\thesection}{S\arabic{section}}
\renewcommand{\thesubsection}{\arabic{subsection}}
\renewcommand{\thesubsubsection}{\arabic{subsubsection}}
\supportingfigtab
\supportingequation

\thispagestyle{empty}
\begin{titlepage}
\centering
\begin{tikzpicture}[remember picture,overlay]
  \fill[abstractbg] (current page.south west) rectangle (current page.north east);
  \node[anchor=north,text=prismslate] at ([yshift=-0.58in]current page.north) {\publicationnote};
  \draw[prismblue,line width=1.1pt] ([xshift=1in,yshift=-1.05in]current page.north west) -- ([xshift=-1in,yshift=-1.05in]current page.north east);
  \draw[prismline,line width=0.7pt] ([xshift=1in,yshift=-1.16in]current page.north west) -- ([xshift=-1in,yshift=-1.16in]current page.north east);
  \draw[prismblue!55,line width=1.0pt] ([xshift=1in,yshift=0.92in]current page.south west) -- ([xshift=0.38\paperwidth,yshift=0.92in]current page.south west);
  \draw[prismline,line width=0.7pt] ([xshift=0.42\paperwidth,yshift=0.92in]current page.south west) -- ([xshift=-1in,yshift=0.92in]current page.south east);
\end{tikzpicture}

\vspace*{0.35cm}
{\sffamily\bfseries\Large\color{prismblue} Supporting Materials}\par
\vspace{0.18cm}
{\sffamily\Large\color{prismslate} for}\par

\vspace{1.35cm}
\begin{tcolorbox}[
  enhanced,
  width=1.06\textwidth,
  height=0.76\textheight,
  valign=center,
  colback=white,
  colframe=prismline,
  boxrule=0.6pt,
  arc=2.5pt,
  left=30pt,
  right=30pt,
  top=30pt,
  bottom=30pt,
  borderline north={1.6pt}{0pt}{prismblue},
  borderline south={0.6pt}{0pt}{prismline}
]
{\bfseries\fontsize{18}{23}\selectfont Bgolearn}\par\vspace{0.25em}
{\bfseries\fontsize{15}{20}\selectfont a unified Bayesian optimization framework for accelerating materials discovery}\par
\vspace{1.0em}
{\color{prismblue}\rule{0.22\linewidth}{1.1pt}}\par
\vspace{0.9em}
{\small Bin Cao$^{1,2}$, Jie Xiong$^{3,4,*}$, Jiaxuan Ma$^{5}$, Yuan Tian$^{3}$, Yirui Hu$^{3}$, Mengwei He$^{6}$, Longhan Zhang$^{1}$, Jiayu Wang$^{7}$, Jian Hui$^{8,*}$, Li Liu$^{7}$, Dezhen Xue$^{9}$, Turab Lookman$^{9,10,*}$, Jun Wang$^{11,*}$, Tong-Yi Zhang$^{1,2,*}$}\par
\vspace{1.0em}
{\scriptsize\color{prismslate}
$^{1}$Guangzhou Municipal Key Laboratory of Materials Informatics, Advanced Materials Thrust, The Hong Kong University of Science and Technology (Guangzhou), Guangzhou, China\par
$^{2}$Department of Physics, City University of Hong Kong, Hong Kong SAR, China\par
$^{3}$Materials Genome Institute, Shanghai University, Shanghai, China\par
$^{4}$State Key Laboratory of Materials for Advanced Nuclear Energy, Shanghai University, Shanghai, China\par
$^{5}$School of Materials Science and Engineering, Shanghai Jiao Tong University, Shanghai, China\par
$^{6}$School of Aerospace, Mechanical and Mechatronic Engineering, School of Computer Science, The University of Sydney, Sydney, NSW, Australia\par
$^{7}$School of Materials Science and Engineering, Harbin Institute of Technology (Shenzhen), Shenzhen, China\par
$^{8}$Suzhou Laboratory, Suzhou, China\par
$^{9}$State Key Laboratory for Mechanical Behavior of Materials, Xi'an Jiaotong University, Xi'an, China\par
$^{10}$AiMaterials Research, Santa Fe, NM, USA\par
$^{11}$University College London, London, UK\par
}\vspace{0.75em}
{\scriptsize\color{prismslate}$^{*}$Corresponding authors: xiongjie@shu.edu.cn; huij@szlab.ac.cn; turablookman@gmail.com; jun.wang@cs.ucl.ac.uk; mezhangt@hkust-gz.edu.cn}
\end{tcolorbox}

\vspace*{0.55cm}
\end{titlepage}

\newpage
\section{Detailed Algorithm Descriptions}
\label{Algorithm}
\textbf{The code booklet provides a tutorial on applying Bgolearn in various situations with version control. Please visit} : \url{https://bgolearn.netlify.app}.

\subsection{Single-Objective Acquisition Functions}

\subsubsection{Expected Improvement (EI)}
\label{eiandvariants}
Expected Improvement is the most widely used acquisition function in Bayesian optimization. For a minimization problem with current best observation $f^* = \min_{i=1}^n y_i$, EI is defined as:

\begin{equation}
\text{EI}(\mathbf{x}) = \mathbb{E}[\max(f^* - f(\mathbf{x}), 0)]
\end{equation}

Under a Gaussian Process surrogate with mean $\mu(\mathbf{x})$ and standard deviation $\sigma(\mathbf{x})$, this has the closed form:

\begin{equation}
\text{EI}(\mathbf{x}) = \begin{cases}
(f^* - \mu(\mathbf{x}))\Phi(Z) + \sigma(\mathbf{x})\phi(Z) & \text{if } \sigma(\mathbf{x}) > 0 \\
0 & \text{if } \sigma(\mathbf{x}) = 0
\end{cases}
\end{equation}

where $Z = (f^* - \mu(\mathbf{x}))/\sigma(\mathbf{x})$, and $\Phi$, $\phi$ are the standard normal CDF and PDF.

\begin{tcolorbox}[colback=codebg, colframe=codeframe,
    boxrule=0.45pt, arc=3pt, left=8pt, right=8pt, top=6pt, bottom=6pt,
    listing only, breakable]
\begin{lstlisting}[style=bgocode]
# Expected Improvement
ei_values, next_point = model.EI()

# With custom baseline
ei_values, next_point = model.EI(T=-2.0)
\end{lstlisting}
\end{tcolorbox}

\textbf{Variants:}
\begin{itemize}
    \item \textbf{EI with Plug-in:} Uses the model prediction on the training data as the baseline ( \texttt{model.EI\_plugin()}).
    \item \textbf{Augmented EI:} Adds an exploration bonus for high-uncertainty regions (\texttt{model.Augmented\_EI()}).
    \item \textbf{Reinterpolation EI:} Accounts for noise in observations (\texttt{model.Reinterpolation\_EI()}).
    \item \textbf{Logarithmic EI:} Applies a logarithmic transform to the EI values to amplify subtle differences (\texttt{model.EI\_log()})).
\end{itemize}

\subsubsection{Upper Confidence Bound (UCB)}

UCB balances exploitation and exploration through an optimistic estimate:

\begin{equation}
\text{UCB}(\mathbf{x}) = \mu(\mathbf{x}) - \beta \sigma(\mathbf{x})
\end{equation}

for minimization (use $\mu(\mathbf{x}) + \beta \sigma(\mathbf{x}$) for maximization). The parameter $\beta$ controls the exploration-exploitation trade-off:
\begin{itemize}
    \item $\beta = 0.5$: Conservative, exploitation-focused
    \item $\beta = 1.0$: Balanced (default)
    \item $\beta = 2.0$: Aggressive exploration
    \item $\beta = 3.0$: Very aggressive, suitable for noisy functions
\end{itemize}

Theoretical analysis suggests $\beta_t = \sqrt{2\log(t^{d/2 + 2}\pi^2/(3\delta))}$ for regret bounds, but practical values are typically 1-3.

\begin{tcolorbox}[colback=codebg, colframe=codeframe,
    boxrule=0.45pt, arc=3pt, left=8pt, right=8pt, top=6pt, bottom=6pt,
    listing only, breakable]
\begin{lstlisting}[style=bgocode]
# UCB with default beta=1.0
ucb_values, next_point = model.UCB()

# Aggressive exploration
ucb_values, next_point = model.UCB(alpha=2.0)
\end{lstlisting}
\end{tcolorbox}

\subsubsection{Probability of Improvement (PI)}

PI computes the probability that a candidate improves upon the current best:

\begin{equation}
\text{PI}(\mathbf{x}) = P(f(\mathbf{x}) < (f^* - \xi)) = \Phi\left(\frac{f^* - \mu(\mathbf{x}) - \xi}{\sigma(\mathbf{x})}\right)
\end{equation}

where $\xi \geq 0$ is an improvement threshold. Setting $\xi > 0$ encourages exploration.

\begin{tcolorbox}[colback=codebg, colframe=codeframe,
    boxrule=0.45pt, arc=3pt, left=8pt, right=8pt, top=6pt, bottom=6pt,
    listing only, breakable]
\begin{lstlisting}[style=bgocode]
# PI with default threshold
poi_values, next_point = model.PoI()

# With exploration threshold
poi_values, next_point = model.PoI(tao=0.01)
\end{lstlisting}
\end{tcolorbox}

\subsubsection{Predictive Entropy Search (PES)}

PES selects points that maximally reduce uncertainty about the location of the optimum:

\begin{equation}
\text{PES}(\mathbf{x}) = H[p(\mathbf{x}^*|\mathcal{D}_n)] - \mathbb{E}_{y}[H[p(\mathbf{x}^*|\mathcal{D}_n \cup \{(\mathbf{x}, y)\})]]
\end{equation}

This is approximated via Monte Carlo sampling from the GP posterior.

\begin{tcolorbox}[colback=codebg, colframe=codeframe,
    boxrule=0.45pt, arc=3pt, left=8pt, right=8pt, top=6pt, bottom=6pt,
    listing only, breakable]
\begin{lstlisting}[style=bgocode]
# PES with 500 MC samples
pes_values, next_point = model.PES(sam_num=500)
\end{lstlisting}
\end{tcolorbox}

\subsubsection{Knowledge Gradient (KG)}

KG estimates the expected value of information from evaluating a point:

\begin{equation}
\text{KG}(\mathbf{x}) = \mathbb{E}[\max_{\mathbf{x}' \in \mathcal{X}} \mu_{n+1}(\mathbf{x}') - \max_{\mathbf{x}' \in \mathcal{X}} \mu_n(\mathbf{x}')]
\end{equation}

where $\mu_{n+1}$ is the posterior mean after observing $(\mathbf{x}, y)$.

\begin{tcolorbox}[colback=codebg, colframe=codeframe,
    boxrule=0.45pt, arc=3pt, left=8pt, right=8pt, top=6pt, bottom=6pt,
    listing only, breakable]
\begin{lstlisting}[style=bgocode]
# Knowledge Gradient
kg_values, next_point = model.Knowledge_G(MC_num=5)
\end{lstlisting}
\end{tcolorbox}

\subsection{Multi-Objective Acquisition Functions}
\label{Acquisition}

\subsubsection{Expected Hypervolume Improvement (EHVI)}

EHVI is the gold standard for multi-objective Bayesian optimization. It maximizes the expected improvement in hypervolume indicator:

\begin{equation}
\text{EHVI}(\mathbf{x}) = \mathbb{E}_{\mathbf{f}(\mathbf{x})}[\max(0, \text{HV}(\mathcal{F} \cup \{\mathbf{f}(\mathbf{x})\}) - \text{HV}(\mathcal{F}))]
\end{equation}

where $\mathcal{F}$ is the current Pareto front and $\text{HV}(\mathcal{F})$ is the hypervolume dominated by $\mathcal{F}$ relative to a reference point $\mathbf{r}$.

\textbf{Hypervolume Computation:}
For a set of points $\mathcal{F} = \{\mathbf{f}_1, \ldots, \mathbf{f}_k\}$ in $m$-dimensional objective space:

\begin{equation}
\text{HV}(\mathcal{F}) = \text{Volume}\left(\bigcup_{i=1}^k [\mathbf{f}_i, \mathbf{r}]\right)
\end{equation}

\textbf{EHVI Calculation:}
Under independent GP surrogates for each objective, EHVI can be computed via:
\begin{enumerate}
    \item Partition objective space into cells based on current Pareto front
    \item For each cell, compute probability that new point falls in cell and dominates
    \item Sum weighted contributions from all cells
\end{enumerate}

For 2D problems, exact computation is feasible. For $m > 2$, we use Monte Carlo approximation:

\begin{equation}
\text{EHVI}(\mathbf{x}) \approx \frac{1}{S}\sum_{s=1}^S \max(0, \text{HV}(\mathcal{F} \cup \{\mathbf{f}^{(s)}(\mathbf{x})\}) - \text{HV}(\mathcal{F}))
\end{equation}

where $\mathbf{f}^{(s)}(\mathbf{x})$ are samples from the GP posterior.

\begin{tcolorbox}[colback=codebg, colframe=codeframe,
    boxrule=0.45pt, arc=3pt, left=8pt, right=8pt, top=6pt, bottom=6pt,
    listing only, breakable]
\begin{lstlisting}[style=bgocode]
from MultiBgolearn import bgo

# EHVI for 3-objective optimization
VS_rec, improvements, idx = bgo.fit(
    'dataset.csv',
    'virtual_space.csv',
    object_num=2,
    method='EHVI',
    assign_model='GaussianProcess',
    bootstrap=8,
    max_search=True  # Maximize all objectives
)
\end{lstlisting}
\end{tcolorbox}

\subsubsection{q-Noisy Expected Hypervolume Improvement (qNEHVI)}

qNEHVI extends EHVI to handle noisy observations and batch acquisition, making it suitable for real-world scenarios with measurement uncertainty and parallel experiments \cite{daulton2021parallel}. It maximizes the expected hypervolume improvement while accounting for observation noise:

\begin{equation}
\text{qNEHVI}(\mathbf{X}_q) = \mathbb{E}_{\mathbf{y} \sim p(\mathbf{y}|\mathcal{D})}[\mathbb{E}_{\mathbf{f}_q \sim p(\mathbf{f}_q|\mathbf{X}_q, \mathcal{D})}[\max(0, \text{HV}(\mathcal{P}(\mathbf{y} \cup \mathbf{f}_q)) - \text{HV}(\mathcal{P}(\mathbf{y})))]]
\end{equation}

where $\mathbf{X}_q = \{\mathbf{x}_1, \ldots, \mathbf{x}_q\}$ is a batch of $q$ candidate points, $\mathcal{D}$ is the observed data, $\mathbf{y}$ represents noisy observations, and $\mathcal{P}(\cdot)$ denotes the Pareto front operator.

\textbf{Key Differences from EHVI:}
\begin{itemize}
    \item \textbf{Noise Modeling:} Explicitly accounts for observation noise $\epsilon \sim \mathcal{N}(0, \sigma_{\text{obs}}^2)$
    \item \textbf{Batch Acquisition:} Selects $q > 1$ points simultaneously for parallel evaluation
    \item \textbf{Noisy Pareto Front:} Computes Pareto front considering uncertainty in existing observations
\end{itemize}

\textbf{Observation Model:}
The noisy observation model is:
\begin{equation}
y_i(\mathbf{x}) = f_i(\mathbf{x}) + \epsilon_i, \quad \epsilon_i \sim \mathcal{N}(0, \sigma_{\text{obs},i}^2)
\end{equation}

where $f_i(\mathbf{x})$ is the true objective value and $\sigma_{\text{obs},i}^2$ is the observation noise variance for objective $i$.

\textbf{Monte Carlo Approximation:}
For computational tractability, qNEHVI is approximated via nested Monte Carlo sampling:

\begin{equation}
\text{qNEHVI}(\mathbf{X}_q) \approx \frac{1}{S_1 S_2}\sum_{s_1=1}^{S_1}\sum_{s_2=1}^{S_2} \max(0, \text{HV}(\mathcal{P}(\mathbf{y}^{(s_1)} \cup \mathbf{f}_q^{(s_2)})) - \text{HV}(\mathcal{P}(\mathbf{y}^{(s_1)})))
\end{equation}

where:
\begin{itemize}
    \item $\mathbf{y}^{(s_1)}$ are samples from noisy observations of existing data
    \item $\mathbf{f}_q^{(s_2)}$ are samples from the GP posterior for candidate batch $\mathbf{X}_q$
    \item Both include observation noise: $\mathbf{y}^{(s)} = \mathbf{f} + \boldsymbol{\epsilon}^{(s)}$
\end{itemize}

\textbf{Batch Selection Strategy:}
For $q > 1$, we employ a greedy sequential approach:
\begin{enumerate}
    \item Initialize batch $\mathbf{X}_q = \emptyset$
    \item For $i = 1$ to $q$:
    \begin{itemize}
        \item Evaluate qNEHVI for all remaining candidates conditioned on $\mathbf{X}_q$
        \item Select $\mathbf{x}_i^* = \arg\max_{\mathbf{x}} \text{qNEHVI}(\mathbf{X}_q \cup \{\mathbf{x}\})$
        \item Update batch: $\mathbf{X}_q \leftarrow \mathbf{X}_q \cup \{\mathbf{x}_i^*\}$
    \end{itemize}
    \item Return $\mathbf{X}_q$
\end{enumerate}

\textbf{Automatic Noise Estimation:}
When observation noise is unknown, it can be estimated from training data:
\begin{equation}
\hat{\sigma}_{\text{obs}} = \alpha \cdot \frac{1}{m}\sum_{i=1}^m \text{std}(\mathbf{y}_i)
\end{equation}

where $\alpha = 0.1$ is a conservative scaling factor and $m$ is the number of objectives.

\begin{tcolorbox}[colback=codebg, colframe=codeframe,
    boxrule=0.45pt, arc=3pt, left=8pt, right=8pt, top=6pt, bottom=6pt,
    listing only, breakable]
\begin{lstlisting}[style=bgocode]
from MultiBgolearn import bgo

# qNEHVI with single point selection
VS_rec, improvements, idx = bgo.fit(
    'dataset.csv',
    'virtual_space.csv',
    object_num=2,
    method='qNEHVI',
    assign_model='RandomForest',
    bootstrap=5,
    batch_size=1,        # Single point
    noise_std=0.05,      # 5% observation noise
    max_search=True
)

# qNEHVI with batch acquisition (parallel experiments)
VS_rec, improvements, idx = bgo.fit(
    'dataset.csv',
    'virtual_space.csv',
    object_num=2,
    method='qNEHVI',
    batch_size=3,        # Select 3 points simultaneously
    noise_std=None,      # Auto-estimate noise
    max_search=True
)


\end{lstlisting}
\end{tcolorbox}

\textbf{When to Use qNEHVI:}
\begin{itemize}
    \item Measurements have significant observation noise or uncertainty
    \item Multiple experiments can be conducted in parallel
    \item Robust optimization under uncertainty is required
    \item Known or estimable observation noise levels
\end{itemize}

\subsubsection{Multi-Objective Probability of Improvement (MO-PI)}

MO-PI extends PI to multiple objectives by computing the probability of dominating at least one point in the current Pareto front:

\begin{equation}
\text{MO-PI}(\mathbf{x}) = P(\exists \mathbf{f}_i \in \mathcal{F} : \mathbf{f}(\mathbf{x}) \prec \mathbf{f}_i)
\end{equation}

where $\prec$ denotes Pareto dominance.

For independent GP surrogates, this is approximated via Monte Carlo:

\begin{equation}
\text{MO-PI}(\mathbf{x}) \approx \frac{1}{S}\sum_{s=1}^S \mathbb{I}\left[\exists \mathbf{f}_i \in \mathcal{F} : \mathbf{f}^{(s)}(\mathbf{x}) \prec \mathbf{f}_i\right]
\end{equation}

\begin{tcolorbox}[colback=codebg, colframe=codeframe,
    boxrule=0.45pt, arc=3pt, left=8pt, right=8pt, top=6pt, bottom=6pt,
    listing only, breakable]
\begin{lstlisting}[style=bgocode]
# MO-PI for bi-objective optimization
VS_rec, improvements, idx = bgo.fit(
    'dataset.csv',
    'virtual_space.csv',
    object_num=2,
    method='PI',
    assign_model='RandomForest',
    bootstrap=5
)
\end{lstlisting}
\end{tcolorbox}

\subsubsection{Multi-Objective Upper Confidence Bound (MO-UCB)}

MO-UCB applies UCB to each objective independently and uses Pareto dominance to select candidates:

\begin{equation}
\text{UCB}_i(\mathbf{x}) = \mu_i(\mathbf{x}) + \beta \sigma_i(\mathbf{x}), \quad i = 1, \ldots, m
\end{equation}

The acquisition value is based on the hypervolume of the UCB vector:

\begin{equation}
\text{MO-UCB}(\mathbf{x}) = \text{HV}(\{\text{UCB}(\mathbf{x})\})
\end{equation}

\begin{tcolorbox}[colback=codebg, colframe=codeframe,
    boxrule=0.45pt, arc=3pt, left=8pt, right=8pt, top=6pt, bottom=6pt,
    listing only, breakable]
\begin{lstlisting}[style=bgocode]
# MO-UCB for 4-objective optimization
VS_rec, improvements, idx = bgo.fit(
    'dataset.csv',
    'virtual_space.csv',
    object_num=2,
    method='UCB',
    assign_model='GaussianProcess',
    bootstrap=10
)
\end{lstlisting}
\end{tcolorbox}

\section{Surrogate Model Details}
\label{Surrogate}
\subsection{Gaussian Process Regression}

\subsubsection{Kernel Functions}

Bgolearn implements several kernel functions:

\textbf{1. Squared Exponential:}
\begin{equation}
k(\mathbf{x}, \mathbf{x}') = \sigma_f^2 \exp\left(-\frac{1}{2}\sum_{d=1}^D \frac{(x_d - x_d')^2}{\ell_d^2}\right)
\end{equation}

\textbf{2. Matérn 5/2:}
\begin{equation}
k(\mathbf{x}, \mathbf{x}') = \sigma_f^2 \left(1 + \frac{\sqrt{5}r}{\ell} + \frac{5r^2}{3\ell^2}\right)\exp\left(-\frac{\sqrt{5}r}{\ell}\right)
\end{equation}
where $r = \|\mathbf{x} - \mathbf{x}'\|_2$.

\textbf{3. Matérn 3/2:}
\begin{equation}
k(\mathbf{x}, \mathbf{x}') = \sigma_f^2 \left(1 + \frac{\sqrt{3}r}{\ell}\right)\exp\left(-\frac{\sqrt{3}r}{\ell}\right)
\end{equation}

\subsubsection{Parameter Optimization}

GP parameters $\boldsymbol{\theta} = \{\sigma_f^2, \ell_1, \ldots, \ell_D, \sigma_n^2\}$ are optimized by maximizing the marginal log-likelihood:

\begin{equation}
\log p(\mathbf{y}|\mathbf{X}, \boldsymbol{\theta}) = -\frac{1}{2}\mathbf{y}^T\mathbf{K}_y^{-1}\mathbf{y} - \frac{1}{2}\log|\mathbf{K}_y| - \frac{n}{2}\log(2\pi)
\end{equation}

where $\mathbf{K}_y = \mathbf{K} + \sigma_n^2\mathbf{I}$ is the covariance matrix with noise.

Bgolearn uses L-BFGS-B optimization with multiple random restarts to avoid local optima.

\subsubsection{Prediction}

For a test point $\mathbf{x}_*$, the posterior distribution is:

\begin{align}
\mu(\mathbf{x}_*) &= \mathbf{k}_*^T \mathbf{K}_y^{-1} \mathbf{y} \\
\sigma^2(\mathbf{x}_*) &= k(\mathbf{x}_*, \mathbf{x}_*) - \mathbf{k}_*^T \mathbf{K}_y^{-1} \mathbf{k}_*
\end{align}

where $\mathbf{k}_* = [k(\mathbf{x}_*, \mathbf{x}_1), \ldots, k(\mathbf{x}_*, \mathbf{x}_n)]^T$.

\subsection{Bootstrap Uncertainty Quantification}

For uncertainty estimation, Bgolearn trains $B$ Random Forest models on bootstrap samples:

\begin{equation}
\mathcal{D}_b = \{(\mathbf{x}_i, y_i) : i \in \text{Bootstrap}(\{1, \ldots, n\})\}, \quad b = 1, \ldots, B
\end{equation}

Predictions from each model $\hat{f}_b(\mathbf{x})$ yield:

\begin{align}
\mu(\mathbf{x}) &= \frac{1}{B}\sum_{b=1}^B \hat{f}_b(\mathbf{x}) \\
\sigma^2(\mathbf{x}) &= \frac{1}{B-1}\sum_{b=1}^B (\hat{f}_b(\mathbf{x}) - \mu(\mathbf{x}))^2
\end{align}

This provides uncertainty estimates comparable to GP at lower computational cost.

\textbf{Recommended Settings:}
\begin{itemize}
    \item Number of trees per forest: 100-500
    \item Bootstrap iterations: 5-10
   
\end{itemize}

\section{Benchmark}
\label{Benchmark_function}

To rigorously evaluate the performance of Bgolearn across diverse optimization scenarios, we selected four canonical benchmark functions that capture the key challenges encountered in materials optimization: high dimensionality, multimodality, and conflicting objectives. These functions are widely used in the optimization literature and provide standardized baselines for comparing algorithmic performance.

\subsection{Single-Objective Benchmarks}

\textit{Hartmann-6D Function.}  
The 6-dimensional Hartmann function is a smooth, multimodal test function with a single global minimum and several local minima. It assesses an optimizer’s ability to explore high-dimensional landscapes without becoming trapped in suboptimal regions. The function is defined as:
\begin{equation}
f_{\text{Hartmann}}(\mathbf{x}) = -\sum_{i=1}^{4} \alpha_i 
\exp\left(-\sum_{j=1}^{6} A_{ij}(x_j - P_{ij})^2\right),
\end{equation}
where $\mathbf{x} \in [0,1]^6$, and the parameters are:
\begin{align*}
\boldsymbol{\alpha} &= [1.0,\, 1.2,\, 3.0,\, 3.2]^T, \\
\mathbf{A} &= 
\begin{bmatrix}
10 & 3 & 17 & 3.5 & 1.7 & 8 \\
0.05 & 10 & 17 & 0.1 & 8 & 14 \\
3 & 3.5 & 1.7 & 10 & 17 & 8 \\
17 & 8 & 0.05 & 10 & 0.1 & 14
\end{bmatrix}, \\
\mathbf{P} &= 10^{-4} \times
\begin{bmatrix}
1312 & 1696 & 5569 & 124 & 8283 & 5886 \\
2329 & 4135 & 8307 & 3736 & 1004 & 9991 \\
2348 & 1451 & 3522 & 2883 & 3047 & 6650 \\
4047 & 8828 & 8732 & 5743 & 1091 & 381
\end{bmatrix}.
\end{align*}
The global minimum is $f^* = -3.32237$ at 
$\mathbf{x}^* = [0.20169,\, 0.15001,\, 0.47687,\, 0.27533,\, 0.31165,\, 0.65730]$.  
This benchmark is particularly relevant to materials optimization, as it mimics composition–property landscapes where multiple local optima exist due to competing phase formations, while a single global optimum represents the best achievable performance.

\textit{Ackley Function.}  
The 5-dimensional Ackley function is highly multimodal, with thousands of local minima, designed to test an optimizer’s exploration capability and resistance to premature convergence. It is defined as:
\begin{equation}
f_{\text{Ackley}}(\mathbf{x}) =
-20 \exp\!\left(-0.2 \sqrt{\frac{1}{D}\sum_{i=1}^{D}x_i^2}\right)
- \exp\!\left(\frac{1}{D}\sum_{i=1}^{D}\cos(2\pi x_i)\right)
+ 20 + e,
\end{equation}
where $\mathbf{x} \in [-5, 5]^5$ and $D = 5$.  
The global minimum is $f^* = 0$ at $\mathbf{x}^* = \mathbf{0}$.  
The function features a nearly flat outer region with numerous cosine-induced local minima surrounding a large central basin. This landscape resembles process–parameter optimization problems in materials science, where minor variations (e.g., temperature or concentration) can create local optima, while the global optimum lies within a narrow parameter regime.

\subsection{Multi-Objective Benchmarks}

\textit{ZDT1 Function.}  
The ZDT1 benchmark is a bi-objective problem with a convex Pareto front, testing an optimizer’s ability to identify and uniformly sample the trade-off surface between two competing objectives. It is defined as:
\begin{align}
f_1(\mathbf{x}) &= x_1, \\
f_2(\mathbf{x}) &= g(\mathbf{x}) \left[1 - \sqrt{\frac{x_1}{g(\mathbf{x})}}\right], \\
g(\mathbf{x}) &= 1 + \frac{9}{D-1}\sum_{i=2}^{D}x_i,
\end{align}
where $\mathbf{x} \in [0,1]^{30}$ and $D = 30$.  
The Pareto-optimal set is characterized by $x_2 = \cdots = x_{30} = 0$, leading to $g(\mathbf{x}^*) = 1$, and the Pareto front is given by:
\[
f_2 = 1 - \sqrt{f_1}, \quad f_1 \in [0,1].
\]
This convex front is analogous to materials design problems where improving one property (e.g., strength) monotonically degrades another (e.g., ductility) following a smooth trade-off relationship.

\textit{DTLZ2 Function.}  
The DTLZ2 benchmark is a tri-objective problem with a spherical Pareto surface, evaluating an optimizer’s ability to handle higher-dimensional objective spaces and discover complex Pareto geometries. It is defined as:
\begin{align}
f_1(\mathbf{x}) &= (1 + g(\mathbf{x}))\cos\!\left(\frac{\pi x_1}{2}\right)\cos\!\left(\frac{\pi x_2}{2}\right), \\
f_2(\mathbf{x}) &= (1 + g(\mathbf{x}))\cos\!\left(\frac{\pi x_1}{2}\right)\sin\!\left(\frac{\pi x_2}{2}\right), \\
f_3(\mathbf{x}) &= (1 + g(\mathbf{x}))\sin\!\left(\frac{\pi x_1}{2}\right), \\
g(\mathbf{x}) &= \sum_{i=3}^{D}(x_i - 0.5)^2,
\end{align}
where $\mathbf{x} \in [0,1]^{12}$ and $D = 12$.  
The Pareto-optimal surface satisfies $x_3 = \cdots = x_{12} = 0.5$, yielding $g(\mathbf{x}^*) = 0$ and a unit sphere in the objective space:
\[
f_1^2 + f_2^2 + f_3^2 = 1.
\]
This spherical geometry represents multi-objective materials design scenarios involving three competing targets (e.g., strength, toughness, and cost), where trade-offs exist in all directions, requiring comprehensive Pareto surface exploration.

\medskip
Together, these four benchmarks span the principal challenges of materials optimization:  
Hartmann-6D tests high-dimensional smooth optimization, Ackley tests multimodal exploration, ZDT1 evaluates bi-objective trade-off discovery, and DTLZ2 assesses tri-objective Pareto surface mapping.  
Performance on these standardized problems provides quantitative evidence of Bgolearn’s capabilities prior to deployment on real materials systems.

\subsection{TPMS designing}
\label{TPMSdesigning}
TPMS structures can be described using an implicit level-set formulation:
\begin{equation}
f(x,y,z) = \alpha_1 f_G(x,y,z) + \alpha_2 f_D(x,y,z),
\end{equation}
subject to the constraint
\begin{equation}
\alpha_1 + \alpha_2 = 1.
\end{equation}

Here, the basis functions $f_G(x,y,z)$ and $f_D(x,y,z)$ correspond to the Gyroid (G) and Diamond (D) minimal surfaces, respectively, and are defined as
\begin{equation}
\begin{aligned}
f_G(x,y,z) &= 
\cos(2\pi x)\sin(2\pi y)
+ \cos(2\pi y)\sin(2\pi z)
+ \cos(2\pi z)\sin(2\pi x)
+ t_1, \\
f_D(x,y,z) &=
\cos(2\pi x)\cos(2\pi y)\cos(2\pi z)
- \sin(2\pi x)\sin(2\pi y)\sin(2\pi z)
+ t_2,
\end{aligned}
\end{equation}
where $t_1$ and $t_2$ are level-set offsets controlling the relative volume fraction of the TPMS structures.

In the above equations, $(x,y,z)$ denote the spatial coordinates, while $\alpha_1$, $\alpha_2$, $t_1$, and $t_2$ are design parameters governing the geometry of the TPMS configuration.

\subsection{Testing settings of High-Entropy Alloys}
\label{Testing_settings_HEA}
Nanoindentation of the newly designed high-entropy alloys (HEAs) was performed using an EVOHT iMicro platform (KLA) in continuous stiffness measurement (CSM) mode. The sample surface was ground to 7000 grit and subsequently polished with 2.5~µm and 0.05~µm suspensions to minimize surface roughness effects. Indents were conducted at a fixed peak load of 50~mN with a loading/unloading rate of 0.1~mN/s and a 5~s dwell at the maximum load. A Poisson’s ratio of 0.33 was used for data reduction.  

\begin{figure}[h]
    \centering
    \includegraphics[width=0.7\columnwidth]{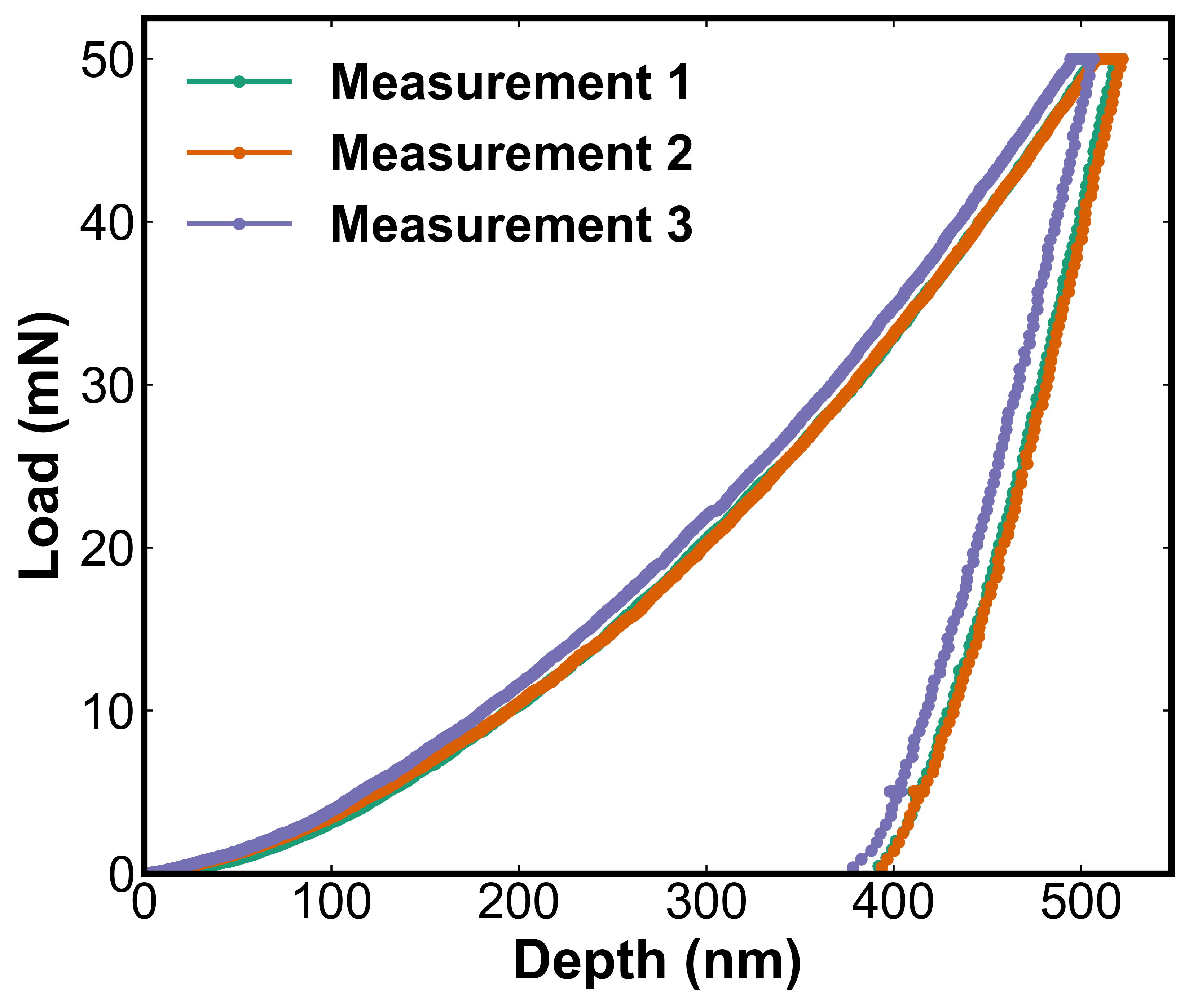}
    \caption{Depth–load curves measured from three independent nanoindentation experiments on the sample. 
    }

    \label{fig:Depth–load}
\end{figure}

Multiple independent indents produced highly overlapping depth-dependent hardness profiles, exhibiting a clear plateau with minimal scatter (Fig.\ref{fig:Depth–load}), demonstrating excellent repeatability across the tested area. Averaging the plateau region yields a hardness of 10.88~±~0.51~GPa, corresponding to 1028.7~±~47.6~HV.

\section{Data}
\label{Data}
\subsection{Triply periodic minimal surface structure}
\centering
\begin{longtable}{ccccc}
\caption{Sampled TPMS configurations and corresponding elastic modulus}
\label{tab:tpms_elasticity} \\
\toprule
$\alpha_1$ & $\alpha_2$ & $t_1$ & $t_2$ & Elastic Modulus (MPa) \\
\midrule
\endfirsthead

\toprule
$\alpha_1$ & $\alpha_2$ & $t_1$ & $t_2$ & Elastic Modulus (MPa) \\
\midrule
\endhead

\midrule
\multicolumn{5}{r}{Continued on next page} \\
\bottomrule
\endfoot

\bottomrule
\endlastfoot

0.814723686 & 0.185276314 & 0.405791937 & -0.373013184 & 1211.254428 \\
0.913375856 & 0.086624144 & 0.132359246 & -0.402459595 & 3110.245652 \\
0.965000000 & 0.035111465 & -0.342000000 & 0.470592782 & 7330.000000 \\
0.957166948 & 0.042833052 & -0.014624351 & 0.300280469 & 3945.858345 \\
0.792000000 & 0.207792670 & 0.459000000 & 0.155740699 & 347.000000 \\
0.679000000 & 0.321264845 & 0.258000000 & 0.243132468 & 250.000000 \\
0.392227020 & 0.607772980 & 0.155477890 & -0.328813312 & 1723.581917 \\
0.706046088 & 0.293953912 & -0.468167154 & -0.223077015 & 7914.498995 \\
0.695000000 & 0.305171377 & -0.183000000 & 0.450222049 & 2430.000000 \\
0.034446081 & 0.965553919 & -0.061255640 & -0.118441543 & 2383.051883 \\
0.765516788 & 0.234483212 & 0.295199901 & -0.313127395 & 1658.187107 \\
0.709000000 & 0.290635169 & 0.255000000 & -0.223974923 & 1560.000000 \\
0.679702677 & 0.320297323 & 0.155098004 & -0.337388265 & 2486.398868 \\
0.340385727 & 0.659614273 & 0.085267751 & -0.276188061 & 1647.683514 \\
0.751267059 & 0.248732941 & -0.244904885 & 0.005957052 & 5098.727230 \\
0.547000000 & 0.452784470 & -0.361000000 & -0.350705994 & 7030.000000 \\
0.257508254 & 0.742491746 & 0.340717256 & -0.245717821 & 1028.237170 \\
0.814284826 & 0.185715174 & -0.256475031 & 0.429263623 & 4614.914175 \\
0.349983766 & 0.650016234 & -0.303404750 & -0.248916142 & 4052.960534 \\
0.616044676 & 0.383955324 & -0.026711151 & -0.148340493 & 2900.128516 \\
0.830828628 & 0.169171372 & 0.085264091 & 0.049723608 & 2671.992089 \\
0.917193664 & 0.082806336 & -0.214160981 & 0.257200229 & 5628.939031 \\
0.753729094 & 0.246270906 & -0.119554153 & 0.067821641 & 3753.298235 \\
0.075854290 & 0.924145710 & -0.446049881 & 0.030797553 & 1064.928484 \\
0.779167230 & 0.220832770 & 0.434010684 & -0.370093792 & 1079.625037 \\
0.568823661 & 0.431176339 & -0.030609359 & -0.488097930 & 4844.727841 \\
0.311000000 & 0.688784958 & 0.028500000 & -0.334351271 & 2910.000000 \\
0.601981941 & 0.398018059 & -0.237028715 & 0.154079098 & 2946.282790 \\
0.689214503 & 0.310785497 & 0.248151593 & -0.049458401 & 1119.755026 \\
0.996000000 & 0.003865283 & -0.422000000 & -0.057321730 & 8560.000000 \\
0.106652770 & 0.893347230 & 0.461898081 & -0.495365776 & 5843.416050 \\
0.774910465 & 0.225089535 & 0.317303221 & 0.368694705 & 552.139576 \\
0.084435846 & 0.915564154 & -0.100217351 & -0.240129597 & 3422.942334 \\
0.800068480 & 0.199931520 & -0.068586173 & 0.410647594 & 2910.196962 \\
0.181847028 & 0.818152972 & -0.236197083 & -0.354461020 & 4950.841991 \\
0.550000000 & 0.450139798 & -0.355000000 & 0.353031118 & 1790.000000 \\
0.622055131 & 0.377944869 & -0.149047619 & 0.013249540 & 3107.607816 \\
0.401808034 & 0.598191966 & -0.424033308 & -0.260083846 & 6019.204325 \\
0.123318935 & 0.876681065 & -0.316092212 & -0.260047474 & 3898.131271 \\
0.945000000 & 0.055212810 & -0.009140000 & -0.010747362 & 3950.000000 \\
0.111000000 & 0.888797245 & 0.280000000 & -0.110261163 & 1480.000000 \\
0.241691286 & 0.758308714 & -0.096087854 & -0.403545475 & 4691.138795 \\
0.575208595 & 0.424791405 & -0.440220457 & -0.265220087 & 7202.226147 \\
0.353158571 & 0.646841429 & 0.321194040 & -0.484596562 & 2605.902790 \\
0.043023802 & 0.956976198 & -0.331009971 & 0.149115475 & 353.857606 \\
0.731722386 & 0.268277614 & 0.147745963 & -0.049076294 & 2000.823277 \\
0.547008892 & 0.452991108 & -0.203679194 & 0.244692807 & 1272.001670 \\
0.188955015 & 0.811044985 & 0.186775433 & -0.316488844 & 2951.653024 \\
0.487000000 & 0.513208368 & -0.064100000 & -0.053216251 & 1220.000000 \\
0.818000000 & 0.182372292 & 0.295000000 & 0.144318130 & 1210.000000 \\

\bottomrule
\end{longtable}

\begin{table}[htbp]
\centering
\caption{TPMS configurations recommended by Bgolearn across optimization iterations}
\label{tab:tpms_high_elasticity}
\begin{tabular}{cccccc}
\toprule
Iteration & $\alpha_1$ & $\alpha_2$ & $t_1$ & $t_2$ & Elastic Modulus (MPa) \\
\midrule
1st & 0.906051736 & 0.093948264 & -0.422657215 & -0.161465350 & 8156.984453 \\
1st & 0.954000000 & 0.045825544 & -0.468000000 & -0.143131014 & 8810.000000 \\
2nd & 0.990381201 & 0.009618799 & -0.466308207 & -0.075746730 & 8914.840133 \\
\rowcolor{gray!20}
2nd & 0.915991244 & 0.084008756 & -0.498848943 & -0.037550841 & \textbf{8945.424651} \\
\bottomrule
\end{tabular}
\end{table}

\subsection{High-entropy alloys}

\begingroup
\setlength{\LTleft}{0.025\textwidth}
\setlength{\LTright}{0.025\textwidth}
\begin{longtable}{c ccccccc c ccccccc}
\caption{
Elemental compositions (at.\%) and Vickers hardness (HV) of
Al--Co--Cr--Cu--Fe--Ni alloys
}
\label{tab:al_co_cr_cu_fe_ni_hv}\\
\toprule
\textbf{ID} & Al & Co & Cr & Cu & Fe & Ni & HV &
\textbf{ID} & Al & Co & Cr & Cu & Fe & Ni & HV \\
\midrule
\endfirsthead

\toprule
\textbf{ID} & Al & Co & Cr & Cu & Fe & Ni & HV &
\textbf{ID} & Al & Co & Cr & Cu & Fe & Ni & HV \\
\midrule
\endhead

\midrule
\multicolumn{16}{r}{Continued on next page} \\
\bottomrule
\endfoot

\bottomrule
\endlastfoot

\cellcolor{gray!20} 1 & 40 & 20 & 20 & 0 & 20 & 0 & 775 & \cellcolor{gray!20} 2 & 40 & 13.3 & 6.7 & 13.3 & 20 & 6.7 & 768 \\
\cellcolor{gray!20} 3 & 43 & 6 & 33 & 6 & 6 & 6 & 764 & \cellcolor{gray!20} 4 & 42.9 & 14.3 & 14.3 & 0 & 14.3 & 14.3 & 740 \\
\cellcolor{gray!20} 5 & 37.5 & 12.5 & 12.5 & 12.5 & 12.5 & 12.5 & 735 & \cellcolor{gray!20} 6 & 25 & 25 & 25 & 0 & 25 & 0 & 720 \\
\cellcolor{gray!20} 7 & 42.9 & 14.3 & 14.3 & 7.1 & 7.1 & 14.3 & 720 & \cellcolor{gray!20} 8 & 25 & 25 & 25 & 0 & 0 & 25 & 712 \\
\cellcolor{gray!20} 9 & 46.2 & 15.4 & 7.7 & 7.7 & 15.4 & 7.7 & 702 & \cellcolor{gray!20} 10 & 42.9 & 21.4 & 7.1 & 7.1 & 7.1 & 14.3 & 701 \\
\cellcolor{gray!20} 11 & 20 & 26.7 & 26.7 & 0 & 26.7 & 0 & 695 & \cellcolor{gray!20} 12 & 38.5 & 15.4 & 15.4 & 0 & 15.4 & 15.4 & 695 \\
\cellcolor{gray!20} 13 & 42.9 & 7.1 & 14.3 & 7.1 & 14.3 & 14.3 & 694 & \cellcolor{gray!20} 14 & 23.8 & 23.8 & 23.8 & 0 & 5 & 23.8 & 665 \\
\cellcolor{gray!20} 15 & 15 & 28.3 & 28.3 & 0 & 28.3 & 0 & 655 & \cellcolor{gray!20} 16 & 35.9 & 12.8 & 12.8 & 12.8 & 12.8 & 12.8 & 655 \\
\cellcolor{gray!20} 17 & 33.3 & 0 & 16.7 & 16.7 & 16.7 & 16.7 & 651 & \cellcolor{gray!20} 18 & 22.2 & 22.2 & 22.2 & 11.1 & 22.2 & 0 & 639 \\
\cellcolor{gray!20} 19 & 33.3 & 13.3 & 13.3 & 13.3 & 13.3 & 13.3 & 625 & \cellcolor{gray!20} 20 & 10 & 30 & 30 & 0 & 30 & 0 & 620 \\
\cellcolor{gray!20} 21 & 16.7 & 16.7 & 33.3 & 0 & 16.7 & 16.7 & 617 & \cellcolor{gray!20} 22 & 23.8 & 23.8 & 23.8 & 0 & 23.8 & 5 & 615 \\
\cellcolor{gray!20} 23 & 30.8 & 15.4 & 15.4 & 7.7 & 15.4 & 15.4 & 609 & \cellcolor{gray!20} 24 & 15.4 & 15.4 & 30.8 & 7.7 & 15.4 & 15.4 & 607 \\
\cellcolor{gray!20} 25 & 20 & 20 & 20 & 10 & 20 & 10 & 604 & \cellcolor{gray!20} 26 & 31.5 & 13.7 & 13.7 & 13.7 & 13.7 & 13.7 & 603 \\
\cellcolor{gray!20} 27 & 25 & 16.7 & 16.7 & 8.3 & 16.7 & 16.7 & 602 & \cellcolor{gray!20} 28 & 16.7 & 16.7 & 25 & 8.3 & 16.7 & 16.7 & 601 \\
\cellcolor{gray!20} 29 & 33.3 & 0 & 13.3 & 13.3 & 13.3 & 26.7 & 593 & \cellcolor{gray!20} 30 & 42.9 & 14.3 & 7.1 & 7.1 & 7.1 & 21.4 & 591 \\
\cellcolor{gray!20} 31 & 22.5 & 22.5 & 22.5 & 0 & 10 & 22.5 & 587 & \cellcolor{gray!20} 32 & 20 & 20 & 20 & 10 & 10 & 20 & 586 \\
\cellcolor{gray!20} 33 & 22.2 & 22.2 & 0 & 11.1 & 22.2 & 22.2 & 584 & \cellcolor{gray!20} 34 & 23 & 15 & 23 & 8 & 15 & 16 & 580 \\
\cellcolor{gray!20} 35 & 30.6 & 0 & 13.9 & 13.9 & 13.9 & 27.8 & 576 & \cellcolor{gray!20} 36 & 28.6 & 14.3 & 14.3 & 14.3 & 14.3 & 14.3 & 576 \\
\cellcolor{gray!20} 37 & 27.3 & 0 & 18.2 & 18.2 & 18.2 & 18.2 & 573 & \cellcolor{gray!20} 38 & 28.6 & 0 & 14.3 & 14.3 & 14.3 & 28.6 & 567 \\
\cellcolor{gray!20} 39 & 21.3 & 21.3 & 21.3 & 0 & 15 & 21.3 & 558 & \cellcolor{gray!20} 40 & 26.5 & 14.7 & 14.7 & 14.7 & 14.7 & 14.7 & 558 \\
\cellcolor{gray!20} 41 & 21.3 & 21.3 & 21.3 & 0 & 21.3 & 15 & 555 & \cellcolor{gray!20} 42 & 20 & 10 & 20 & 10 & 20 & 20 & 551 \\
\cellcolor{gray!20} 43 & 23.8 & 5 & 23.8 & 0 & 23.8 & 23.8 & 550 & \cellcolor{gray!20} 44 & 24.2 & 0 & 15.2 & 15.2 & 15.2 & 30.3 & 550 \\
\cellcolor{gray!20} 45 & 22.5 & 22.5 & 22.5 & 0 & 22.5 & 10 & 548 & \cellcolor{gray!20} 46 & 23.1 & 0 & 15.4 & 15.4 & 15.4 & 30.8 & 546 \\
\cellcolor{gray!20} 47 & 20 & 20 & 10 & 10 & 20 & 20 & 546 & \cellcolor{gray!20} 48 & 22.2 & 11.1 & 0 & 22.2 & 22.2 & 22.2 & 545 \\
\cellcolor{gray!20} 49 & 26.5 & 0 & 14.7 & 14.7 & 14.7 & 29.4 & 544 & \cellcolor{gray!20} 50 & 22.5 & 10 & 22.5 & 0 & 22.5 & 22.5 & 539 \\
\cellcolor{gray!20} 51 & 17.9 & 20.5 & 20.5 & 0 & 20.5 & 20.5 & 538 & \cellcolor{gray!20} 52 & 16.7 & 16.7 & 16.7 & 8.3 & 25 & 16.7 & 537 \\
\cellcolor{gray!20} 53 & 20 & 20 & 0 & 20 & 20 & 20 & 536 & \cellcolor{gray!20} 54 & 22.2 & 0 & 22.2 & 11.1 & 22.2 & 22.2 & 534 \\
\cellcolor{gray!20} 55 & 16.7 & 33.3 & 16.7 & 0 & 16.7 & 16.7 & 532 & \cellcolor{gray!20} 56 & 23.8 & 4.8 & 0 & 23.8 & 23.8 & 23.8 & 531 \\
\cellcolor{gray!20} 57 & 18.4 & 20.4 & 20.4 & 0 & 20.4 & 20.4 & 527 & \cellcolor{gray!20} 58 & 19.4 & 0 & 16.1 & 16.1 & 16.1 & 32.3 & 521 \\
\cellcolor{gray!20} 59 & 20 & 20 & 20 & 0 & 20 & 20 & 516 & \cellcolor{gray!20} 60 & 15.4 & 15.4 & 15.4 & 7.7 & 30.8 & 15.4 & 514 \\
\cellcolor{gray!20} 61 & 21.3 & 21.3 & 15 & 0 & 21.3 & 21.3 & 510 & \cellcolor{gray!20} 62 & 16.7 & 16.7 & 16.7 & 0 & 33.3 & 16.7 & 510 \\
\cellcolor{gray!20} 63 & 23.1 & 15.4 & 15.4 & 15.4 & 15.4 & 15.4 & 510 & \cellcolor{gray!20} 64 & 18.2 & 18.2 & 18.2 & 0 & 18.2 & 27.3 & 503 \\
\cellcolor{gray!20} 65 & 11 & 0 & 29 & 29 & 5 & 26 & 495 & \cellcolor{gray!20} 66 & 21.7 & 0 & 21.7 & 21.7 & 21.7 & 13 & 494 \\
\cellcolor{gray!20} 67 & 23.8 & 19 & 19 & 0 & 19 & 19 & 487 & \cellcolor{gray!20} 68 & 20.8 & 0 & 20.8 & 20.8 & 20.8 & 16.7 & 486 \\
\cellcolor{gray!20} 69 & 12 & 0 & 31 & 20 & 5 & 32 & 483 & \cellcolor{gray!20} 70 & 31 & 17.2 & 17.2 & 0 & 17.2 & 17.2 & 482 \\
\cellcolor{gray!20} 71 & 23.1 & 19.2 & 19.2 & 0 & 19.2 & 19.2 & 479 & \cellcolor{gray!20} 72 & 20 & 0 & 20 & 20 & 20 & 20 & 479 \\
\cellcolor{gray!20} 73 & 11 & 0 & 28 & 29 & 7 & 25 & 477 & \cellcolor{gray!20} 74 & 22.5 & 22.5 & 10 & 0 & 22.5 & 22.5 & 476 \\
\cellcolor{gray!20} 75 & 5 & 31.7 & 31.7 & 0 & 31.7 & 0 & 475 & \cellcolor{gray!20} 76 & 20.6 & 15.9 & 15.9 & 15.9 & 15.9 & 15.9 & 475 \\
\cellcolor{gray!20} 77 & 18.2 & 9.1 & 18.2 & 18.2 & 18.2 & 18.2 & 473 & \cellcolor{gray!20} 78 & 10 & 0 & 35 & 25 & 5 & 25 & 472 \\
\cellcolor{gray!20} 79 & 12 & 0 & 31 & 21 & 5 & 31 & 469 & \cellcolor{gray!20} 80 & 11 & 0 & 28 & 27 & 6 & 28 & 469 \\
\cellcolor{gray!20} 81 & 13 & 0 & 28 & 22 & 6 & 31 & 459 & \cellcolor{gray!20} 82 & 10 & 0 & 35 & 26 & 5 & 24 & 454 \\
\cellcolor{gray!20} 83 & 16.7 & 25 & 16.7 & 8.3 & 16.7 & 16.7 & 451 & \cellcolor{gray!20} 84 & 11.8 & 0 & 29.4 & 0 & 44.1 & 14.7 & 450 \\
\cellcolor{gray!20} 85 & 10 & 0 & 35 & 24 & 5 & 26 & 441 & \cellcolor{gray!20} 86 & 23.8 & 23.8 & 5 & 0 & 23.8 & 23.8 & 438 \\
\cellcolor{gray!20} 87 & 25 & 25 & 0 & 0 & 25 & 25 & 430 & \cellcolor{gray!20} 88 & 18.2 & 18.2 & 18.2 & 18.2 & 18.2 & 9.1 & 423 \\
\cellcolor{gray!20} 89 & 18.2 & 18.2 & 18.2 & 18.2 & 9.1 & 18.2 & 418 & \cellcolor{gray!20} 90 & 20 & 20 & 20 & 20 & 0 & 20 & 415 \\
\cellcolor{gray!20} 91 & 19.2 & 0 & 19.2 & 19.2 & 19.2 & 23.1 & 408 & \cellcolor{gray!20} 92 & 16.7 & 0 & 16.7 & 16.7 & 16.7 & 33.3 & 392 \\
\cellcolor{gray!20} 93 & 15.8 & 21.1 & 21.1 & 0 & 21.1 & 21.1 & 388 & \cellcolor{gray!20} 94 & 16.7 & 20.8 & 20.8 & 0 & 20.8 & 20.8 & 382 \\
\cellcolor{gray!20} 95 & 11.1 & 0 & 22.2 & 22.2 & 22.2 & 22.2 & 382 & \cellcolor{gray!20} 96 & 20 & 26.7 & 0 & 0 & 26.7 & 26.7 & 380 \\
\cellcolor{gray!20} 97 & 16.7 & 0 & 55.6 & 0 & 0 & 27.8 & 371 & \cellcolor{gray!20} 98 & 18.5 & 0 & 18.5 & 18.5 & 18.5 & 25.9 & 370 \\
\cellcolor{gray!20} 99 & 18.2 & 18.2 & 9.1 & 18.2 & 18.2 & 18.2 & 367 & \cellcolor{gray!20} 100 & 18.2 & 27.3 & 0 & 18.2 & 18.2 & 18.2 & 366 \\
\cellcolor{gray!20} 101 & 16.7 & 16.7 & 16.7 & 8.3 & 16.7 & 25 & 358 & \cellcolor{gray!20} 102 & 15.3 & 0 & 16.9 & 16.9 & 16.9 & 33.9 & 339 \\
\cellcolor{gray!20} 103 & 14.9 & 21.3 & 21.3 & 0 & 21.3 & 21.3 & 338 & \cellcolor{gray!20} 104 & 16.7 & 16.7 & 16.7 & 0 & 16.7 & 33.3 & 335 \\
\cellcolor{gray!20} 105 & 13.8 & 0 & 17.2 & 17.2 & 17.2 & 34.5 & 315 & \cellcolor{gray!20} 106 & 15.4 & 15.4 & 15.4 & 7.7 & 15.4 & 30.8 & 310 \\
\cellcolor{gray!20} 107 & 6.3 & 0 & 31.3 & 0 & 46.9 & 15.6 & 304 & \cellcolor{gray!20} 108 & 15.4 & 30.8 & 15.4 & 7.7 & 15.4 & 15.4 & 295 \\
\cellcolor{gray!20} 109 & 12.3 & 0 & 17.5 & 17.5 & 17.5 & 35.1 & 290 & \cellcolor{gray!20} 110 & 8 & 17 & 17 & 8 & 17 & 33 & 280 \\
\cellcolor{gray!20} 111 & 10.7 & 0 & 17.9 & 17.9 & 17.9 & 35.7 & 278 & \cellcolor{gray!20} 112 & 13.8 & 17.2 & 17.2 & 17.2 & 17.2 & 17.2 & 273 \\
\cellcolor{gray!20} 113 & 15.4 & 15.4 & 15.4 & 0 & 15.4 & 38.5 & 265 & \cellcolor{gray!20} 114 & 16.7 & 33.3 & 0 & 16.7 & 16.7 & 16.7 & 249 \\
\cellcolor{gray!20} 115 & 14.3 & 14.3 & 14.3 & 0 & 14.3 & 42.9 & 242 & \cellcolor{gray!20} 116 & 9.1 & 0 & 18.2 & 18.2 & 18.2 & 36.4 & 238 \\
\cellcolor{gray!20} 117 & 11.1 & 0 & 22.2 & 0 & 22.2 & 44.4 & 229 & \cellcolor{gray!20} 118 & 12.5 & 12.5 & 12.5 & 0 & 12.5 & 50 & 225 \\
\cellcolor{gray!20} 119 & 9.1 & 18.2 & 18.2 & 18.2 & 18.2 & 18.2 & 207 & \cellcolor{gray!20} 120 & 14.3 & 28.6 & 0 & 0 & 28.6 & 28.6 & 205 \\
\cellcolor{gray!20} 121 & 7.4 & 0 & 18.5 & 18.5 & 18.5 & 37 & 200 & \cellcolor{gray!20} 122 & 5.7 & 18.9 & 18.9 & 18.9 & 18.9 & 18.9 & 185 \\
\cellcolor{gray!20} 123 & 0 & 25 & 25 & 25 & 0 & 25 & 183 & \cellcolor{gray!20} 124 & 0 & 23.8 & 23.8 & 23.8 & 5 & 23.8 & 182 \\
\cellcolor{gray!20} 125 & 0 & 33.3 & 16.7 & 16.7 & 16.7 & 16.7 & 175 & \cellcolor{gray!20} 126 & 0 & 16.7 & 33.3 & 16.7 & 16.7 & 16.7 & 172 \\
\cellcolor{gray!20} 127 & 0 & 22.5 & 22.5 & 22.5 & 10 & 22.5 & 172 & \cellcolor{gray!20} 128 & 0 & 21.3 & 21.3 & 21.3 & 15 & 21.3 & 171 \\
\cellcolor{gray!20} 129 & 0 & 22.5 & 22.5 & 22.5 & 22.5 & 10 & 170 & \cellcolor{gray!20} 130 & 0 & 28.3 & 28.3 & 0 & 28.3 & 15 & 170 \\
\cellcolor{gray!20} 131 & 5.7 & 0 & 18.9 & 18.9 & 18.9 & 37.7 & 170 & \cellcolor{gray!20} 132 & 5.3 & 21.1 & 21.1 & 0 & 26.3 & 26.3 & 168 \\
\cellcolor{gray!20} 133 & 0 & 21.3 & 21.3 & 21.3 & 21.3 & 15 & 167 & \cellcolor{gray!20} 134 & 14.3 & 42.9 & 0 & 14.3 & 14.3 & 14.3 & 166 \\
\cellcolor{gray!20} 135 & 3.8 & 0 & 19.2 & 19.2 & 19.2 & 38.5 & 162 & \cellcolor{gray!20} 136 & 0 & 21.3 & 15 & 21.3 & 21.3 & 21.3 & 161 \\
\cellcolor{gray!20} 137 & 0 & 15 & 21.3 & 21.3 & 21.3 & 21.3 & 158 & \cellcolor{gray!20} 138 & 0 & 22.5 & 10 & 22.5 & 22.5 & 22.5 & 158 \\
\cellcolor{gray!20} 139 & 0 & 16.7 & 16.7 & 16.7 & 16.7 & 33.3 & 158 & \cellcolor{gray!20} 140 & 0 & 16.7 & 16.7 & 16.7 & 33.3 & 16.7 & 157 \\
\cellcolor{gray!20} 141 & 0 & 20 & 20 & 20 & 20 & 20 & 155 & \cellcolor{gray!20} 142 & 0 & 25 & 0 & 25 & 25 & 25 & 154 \\
\cellcolor{gray!20} 143 & 0 & 23.8 & 5 & 23.8 & 23.8 & 23.8 & 153 & \cellcolor{gray!20} 144 & 0 & 10 & 22.5 & 22.5 & 22.5 & 22.5 & 150 \\
\cellcolor{gray!20} 145 & 7 & 0 & 23.3 & 0 & 23.3 & 46.5 & 149 & \cellcolor{gray!20} 146 & 0 & 5 & 23.8 & 23.8 & 23.8 & 23.8 & 146 \\
\cellcolor{gray!20} 147 & 0 & 0 & 25 & 25 & 25 & 25 & 143 & \cellcolor{gray!20} 148 & 0 & 26.7 & 26.7 & 0 & 26.7 & 20 & 140 \\
\cellcolor{gray!20} 149 & 0 & 25 & 25 & 0 & 25 & 25 & 139 & \cellcolor{gray!20} 150 & 7.7 & 30.8 & 0 & 0 & 30.8 & 30.8 & 135 \\
\cellcolor{gray!20} 151 & 8.6 & 22.9 & 22.9 & 0 & 22.9 & 22.9 & 131 & \cellcolor{gray!20} 152 & 9.1 & 22.7 & 22.7 & 0 & 22.7 & 22.7 & 127 \\
\cellcolor{gray!20} 153 & 0 & 20 & 20 & 0 & 20 & 40 & 125 & \cellcolor{gray!20} 154 & 2.4 & 24.4 & 24.4 & 0 & 24.4 & 24.4 & 118 \\
\cellcolor{gray!20} 155 & 5.9 & 23.5 & 23.5 & 0 & 23.5 & 23.5 & 110 & & & & & & & & \\
\end{longtable}

\newpage

\subsection{Medium-manganese steels}

\begin{table}[htbp]
\centering
\caption{Ultimate yield strength (YS, MPa) and total elongation (EL, \%) of Fe–0.3C–8Mn–2Al across 16 heat-treatment conditions.}
\label{tab:heat_treatment_mechanical_properties}
\resizebox{\textwidth}{!}{%
\begin{tabular}{ccccc}
\toprule
AustTemp ($^\circ$C) &
AnnTemp ($^\circ$C) &
AnnTime (min) &
YS (MPa) &
TE (\%) \\
\midrule
700$ \pm$ 2 & 610$ \pm$ 2 & 30$ \pm$ 2  & 702$ \pm$ 14 & 27.4$ \pm$ 1.6 \\
700$ \pm$ 2 & 640$ \pm$ 2 & 60$ \pm$ 2  & 653$ \pm$ 20 & 18.4$ \pm$ 3.9 \\
700$ \pm$ 2 & 670$ \pm$ 2 & 90$ \pm$ 2  & 655$ \pm$ 7  & 26.9$ \pm$ 0.8 \\
700$ \pm$ 2 & 700$ \pm$ 2 & 120$ \pm$ 2 & 504$ \pm$ 7  & 11.6$ \pm$ 0.5 \\
760$ \pm$ 2 & 610$ \pm$ 2 & 60$ \pm$ 2  & 781$ \pm$ 10 & 31.7$ \pm$ 2.8 \\
760$ \pm$ 2 & 640$ \pm$ 2 & 30$ \pm$ 2  & 777$ \pm$ 14 & 46.6$ \pm$ 1.7 \\
760$ \pm$ 2 & 670$ \pm$ 2 & 120$ \pm$ 2 & 675$ \pm$ 13 & 39.1$ \pm$ 5.3 \\
760$ \pm$ 2 & 700$ \pm$ 2 & 90$ \pm$ 2  & 582$ \pm$ 14 & 17.7$ \pm$ 2.5 \\
820$ \pm$ 2 & 610$ \pm$ 2 & 90$ \pm$ 2  & 782$ \pm$ 13 & 30.9$ \pm$ 2.1 \\
820$ \pm$ 2 & 640$ \pm$ 2 & 120$ \pm$ 2 & 722$ \pm$ 1  & 51.2$ \pm$ 1.2 \\
820$ \pm$ 2 & 670$ \pm$ 2 & 30$ \pm$ 2  & 694$ \pm$ 2  & 61.5$ \pm$ 3.3 \\
820$ \pm$ 2 & 700$ \pm$ 2 & 60$ \pm$ 2  & 619$ \pm$ 1  & 35.6$ \pm$ 4.2 \\
880$ \pm$ 2 & 610$ \pm$ 2 & 120$ \pm$ 2 & 702$ \pm$ 8  & 34.2$ \pm$ 0.9 \\
880$ \pm$ 2 & 640$ \pm$ 2 & 90$ \pm$ 2  & 663$ \pm$ 9  & 50.9$ \pm$ 2.3 \\
880$ \pm$ 2 & 670$ \pm$ 2 & 60$ \pm$ 2  & 531$ \pm$ 9  & 57.1$ \pm$ 4.3 \\
880$ \pm$ 2 & 700$ \pm$ 2 & 30$ \pm$ 2  & 606$ \pm$ 12 & 48.5$ \pm$ 1.7 \\
\bottomrule
\end{tabular}}
\end{table}

\end{document}